\documentclass[]{interact}
\usepackage{epstopdf}
\usepackage[caption=false]{subfig}
\usepackage[numbers,sort&compress]{natbib}
\usepackage{epsfig}
\bibpunct[, ]{[}{]}{,}{n}{,}{,}

\makeatletter
\def\NAT@def@citea{\def\@citea{\NAT@separator}}
\makeatother
\theoremstyle{plain}

\theoremstyle{definition}

\theoremstyle{remark}

\begin{document}

\articletype{Review}

\title{Quantum spin liquids}

\author{
\name{Tom Lancaster\textsuperscript{a}\thanks{CONTACT Tom Lancaster. Email: tom.lancaster@durham.ac.uk}}
\affil{\textsuperscript{a}Department of Physics, Durham University,
  South Road, Durham, DH1 3LE, UK}
}
\maketitle

\begin{abstract}
  A glance at recent research on magnetism turns up a curious set of
articles discussing, or claiming evidence for, a
state of matter called a quantum spin liquid (QSL). These articles are
notable in their invocation of exotic notions of topological physics, quantum entanglement, fractional
quantum numbers, anyon statistics and gauge field theories.
So what is a QSL and why do we need this complicated technical
vocabulary to describe it?
Our aim in
this article is to introduce some of these concepts and provide a discussion
of what a QSL is, where it might occur in Nature and why it is
of interest. As we'll
see, this is a rich subject which is still in development, and
unambiguous evidence for the realisation of the QSL state in a
magnetic material 
remains hotly debated. However, the payoff in terms of
the
special nature of quantum entanglement in the
QSL, and its diverse spectrum of unusual excitations and 
topological status will (at least to some extent)
justify the need to engage with some powerful, occasionally abstract, technical material. 
\end{abstract}


\section{Introduction}

We start with an attempt at a working
definition of a quantum spin liquid (QSL) taken
from Ref.~\cite{wen}:
{\it a quantum spin liquid ground state is an electronic insulator with spin-rotation symmetry and an odd
number of electrons per unit cell.} Although there's quite a lot going on
here, one key feature is that
we have an interacting electronic system comprising localised
electrons whose spin moments do
not align into a magnetically-ordered configuration, even down to
$T=0$.
Although this notion of a lack of spin order will provide a jumping-off point, we'll see
that the picture of a disordered magnetic state fails to capture much
of the rich physics that underlies the QSL. However, the general principle
of order will underly our discussion. 
Specifically, we shall start from Lev Landau's description of the
order that follows from
a symmetry-breaking phase transition \cite{landau}, and the subsequently development
of this subject by Philip
Anderson \cite{anderson}, since  these concepts have provided a means
of classifying and understanding ordered phases of
condensed matter. 

The discovery and elucidation of the fractional quantum Hall (FQH)
fluid fundamentally  challenged this pervasive notion of order \cite{wen,lancaster}.
The FQH fluid is a state of matter realised in a two-dimensional
electron system in high magnetic fields \cite{singleton}. 
The fluid has a gap in energy between the ground state and the
lowest-lying excited
states of the system.\footnote{The existence or absence of such energy gaps will turn out to be a useful
notion in classifying states of matter.}
If we put enough energy into the system to excite excitations
over the energy gap, the resulting particle-like states do not carry
integer numbers of elementary electronic charges, but rather
fractions of an electron charge. We say that they are
{\it fractionalised}. Moreover, they are neither bosons nor fermions, but
are known as anyons, and have  their own, unique, quantum statistics. 
There is not one FQH fluid, but many, each realised in practice
by applying a magnetic field of a different magnitude. Moreover, we cannot
distinguish or classify the different FQH states
by their symmetries, as we can with other ordered states of matter. 
Instead,  topology, the
study of the shapes of objects and spaces, is invoked,
and we 
treat the FQH fluid as being {\it topologically ordered}
(TO). Relative to conventionally-ordered states, TO
states are typified by (i) a
 ground-state degeneracy that depends on the topology of the system, (ii) a gap against excitations
and (iii) a high degree of quantum entanglement. 

Here the concept of a QSL fits into condensed matter. Like the FQH
fluids, 
there is not one QSL state, but many that can't be distinguished by symmetry.
All QSLs have fractionalised excitations. 
Some QSLs show topological order and have gapped excitations; some do
not.  QSLs
feature an unusually-high degree of
(long-range)
quantum entanglement in their ground
states. We will take the view that it is the 
nature of the quantum
entanglement that characterises the QSL and explains its 
properties, rather than its magnetic disorder. Following X.-G.\ Wen \cite{wen}, we 
call the special entanglement {\it quantum order} as a shorthand,\footnote{The entanglement partially explains the Q in QSL. The other
  feature is that we are dealing with $T=0$, where fluctuations are
  purely quantum mechanical rather than thermal. In contrast, a
  classical spin liquid has $T=0$ order destroyed by $T>0$ thermal fluctuations
owing to high ground-state degeneracy.} 
because the unique pattern of
entanglements  characterises each QSL. 
Topological order is then a special case of
quantum order in which all excitations have finite
energy gaps between the ground state and the excited states.\footnote{To close the loop on our technical definition above,
  with its specification of odd numbers of electrons,  we
  note that the 
  {\it Lieb-Schultz-Mattis} theorem says that a system with
  a half-odd integer spin per unit cell, and without symmetry breaking order, has either
  (i) a unique ground state with gapless excitations (e.g.\ a gapless QSL) or (ii) a
  degenerate ground state with a gap (e.g.\ a topologically ordered QSL).}

The paper is structured as follows: we start with an introduction to
the treatment of order and its absence in magnets, followed in
Section~\ref{sec:ingredients}
by a survey of the conceptual ingredients required to understand this
field. Two of these: anyons and gauge theory are introduced in more
detail in Sections~\ref{sec:anyons} and \ref{sec:gauge}. We then
discuss the toric code, which is the simplest model of a
QSL. Fractionalised spinons are described in Section~\ref{sec:spinons}
as a route to finding more model QSLs, before the exactly solvable
Kitaev model is introduced in Section~\ref{sec:kitaev}. Finally we
briefly assess some examples in  the search for a realisation of the QSL state.

\section{Background: magnetic order, disorder and the RVB}\label{sec:bg}
\begin{figure}
  \begin{center}
    \epsfig{file=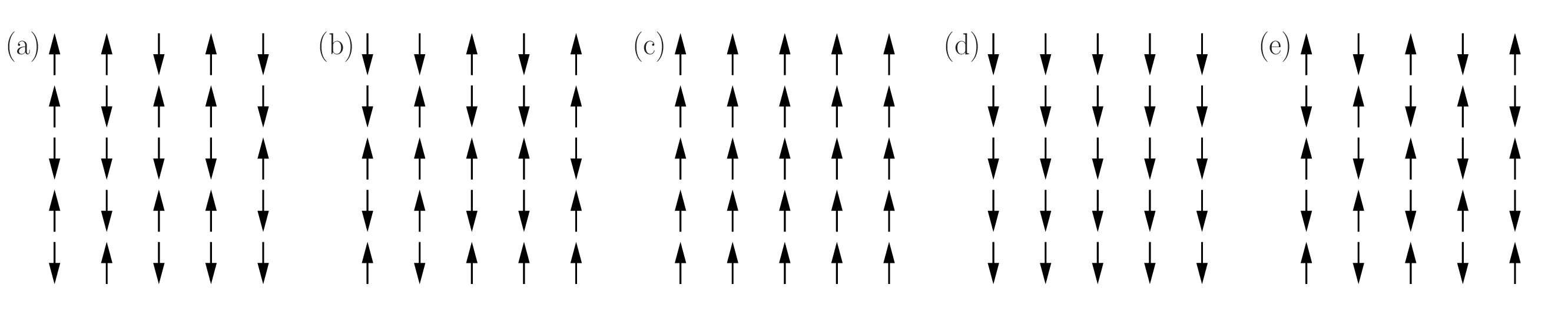,width=\columnwidth}
\caption{(a) A disordered (paramagnetic) magnetic state. This configuration has magnetisation
  $M=0$ and so does (b), which is produced by turning all of the
  spins through 180$^{\circ}$. (c) An ordered (ferro)magnetic phase
  with $M=M_{0}$. Turning the spins through 180$^{\circ}$ results in
  (d) which has $M=-M_{0}.$ (e) The N\'{e}el antiferromagnet, with
  ordered, alternating spins.\label{fig:magnet1}}
\end{center}
\end{figure}

A simple picture of a magnetically disordered system is shown in
Fig.~\ref{fig:magnet1}(a).
The arrows represent spins
arranged on an ordered lattice. (This structural order is always
assumed, the disorder refers only to the arrangement of magnetic
moment directions.)
The system can be understood \cite{blundell,anderson,landau} via its symmetry:
the magnetisation $M$ (that is, the average magnetic moment) is zero, since as many arrows
point up as down. If we turn each of the
moments through 180$^{\circ}$ then we obtain the situation shown in Fig.~\ref{fig:magnet1}(b), which also has
$M=0$. We therefore cannot tell from the magnetisation that we have transformed
the system. This inability to tell that a change has been made is a symmetry. 
An ordered magnet is shown in
Fig.~\ref{fig:magnet1}(c).
This has magnetisation $M_{0}\neq 0$ and 
has lost its previous symmetry: turning the arrows
through 180$^{\circ}$ [Fig.~\ref{fig:magnet1}(d)]
makes a measurable difference in that it
reverses  the magnetisation $M_{0}\rightarrow
-M_{0}$. The symmetry has been broken on magnetic ordering.
In Nature, this sort of ordering is observed to
take place via a magnetic phase transition at a temperature $T_{\mathrm{c}}$.

Philip Anderson notes some key points here \cite{anderson} that apply
in general to broken-symmetry states. 
(i) Once the rotational symmetry of the spin system is broken,
the ordered magnetic state becomes rigid, such that it costs energy
to deform the spin structure. (ii) A new sort of excitation emerges on
symmetry breaking: the magnon. This can be thought of as a single
flipped spin, smeared out to give spin $s=1$
particle-like excitation.
Owing to the possibility of exciting magnons with arbitrarily long
wavelength, for a rotationally-invariant Hamiltonian it costs a vanishing amount of energy to
create one, and so, formally, we say that the excitation is gapless. 
One way to think about the gaplessness is that it is ``protected'' by
symmetry and the ordering.
 
Allowing spins now to point in any direction, a magnet has a microscopic description using the
Heisenberg model with Hamiltonian \cite{blundell}
$\hat{H} = -\sum_{\langle ij\rangle} J_{ij} \hat{\boldsymbol{S}}_{i}\cdot \hat{\boldsymbol{S}}.$
We will consider only $S=1/2$ spin operators $\hat{\boldsymbol{S}}$ and
the sum will be taken only over nearest-neighbour spins, denoted
$\langle ij \rangle$. The
ferromagnet (FM) [Fig.~\ref{fig:magnet1}(c)] is found for constant exchange $J>0$, but
more common than the ferromagnet is the antiferromagnet (AFM) ($J<0$),
where magnetic sublattices break symmetry, with the spins arranged
into an alternating N\'{e}el state [Fig.~\ref{fig:magnet1}(e)]. 

\begin{figure}
  \begin{center}
    \epsfig{file=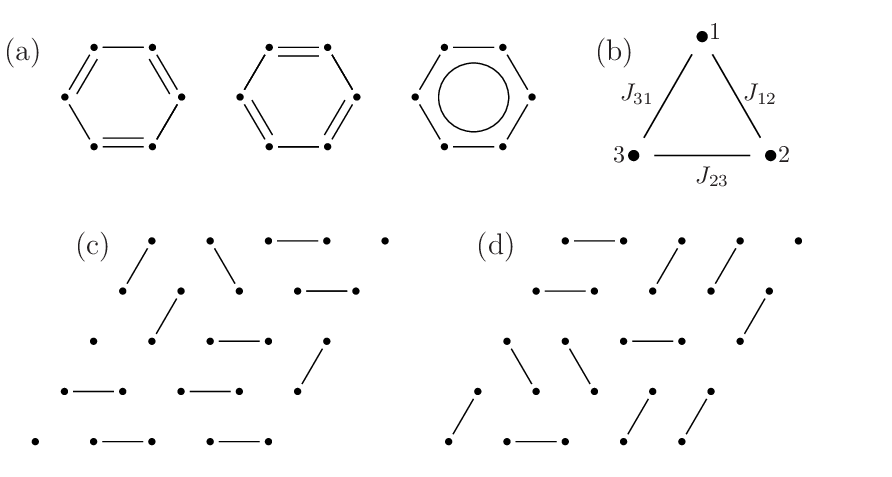,width=10cm}\\
    \caption{(a) Benzene has bonds that we picture in the $\phi_{1}$
       (left) or $\phi_{2}$ (middle) states. The superposition of
       these leads to the $\pi$ orbital picture (right). 
      (b) A triangular arrangement of three spins with exchange
      interactions between them. (c,d) Two possible RVB configurations
      of singlet bonds linking spins on the triangular lattice.\label{fig:benz}}
\end{center}
\end{figure}

Antiferromagnets have several differences compared to ferromagnets, especially when
quantum mechanics is taken into account \cite{blundell,anderson}. The crux is that the
N\'{e}el state is not an eigenstate of the
Heisenberg Hamiltonian.
It's useful at this stage to take a brief step back and consider the
quantum-mechanical fate of a model two-level system such as the benzene
molecule \cite{feynman}.
Benzene has two energetically-degenerate configurations of alternating double and
single covalent bonds, as shown in Fig.~\ref{fig:benz}(a)
(states we shall call $\phi_{1}$ and $\phi_{2}$). Neither is an eigenstate
of the system, and there is a
matrix element for a transition between them. 
It is straightforward to show \cite{feynman} that the ground state 
is the symmetric superposition of the states
[($\phi_{1}+\phi_{2}$)/$\sqrt{2}$], and
 the antisymmetric combination
[($\phi_{1}-\phi_{2}$)/$\sqrt{2}$] is an excited state.
The ground state superposition gives rise to the notion of a
delocalised $\pi$ orbital [Fig~\ref{fig:benz}(a, right)] that represents the system fluctuating, or resonating,
between the two alternating-bond configurations. 
Specifically, if we start the system in state
$\phi_{1}$, then the probability of finding the
system in this same state some time $t$ later oscillates as a function of $t$, at a
frequency $\omega = 2V/\hbar$, determined by the energy gap $2V$
between the eigenstates. These fluctuations have nothing to do with
temperature, but are purely quantum-mechanical in nature. The
point here is that since the initial state $\phi_{1}$ is not an eigenstate of the
system, probability sloshes around at a rate determined by the
energy-level separation.
Clearly, if $V$ is small, the resulting small energy-level
separation means that the oscillations have correspondingly low
frequency. If it approaches zero, the state $\phi_{1}$ becomes stable,
even though it is not an eigenstate.

Since, like $\phi_{1,2}$, the N\'{e}el state is not an eigenstate of the
AFM Hamiltonian, it  was suggested that quantum
fluctuations (generalising the sloshing of the time-dependence of the probability density in the previous example) would break up an
antiferromagnetically-ordered configuration of spins. 
However, the N\'{e}el state {\it is} stable. This is because, for the
large systems that comprise condensed matter,
there are lots of eigenstates arbitrarily close in energy  out of which a
stable wavepacket can be constructed. The
closeness of eigenstates in energy means that the
oscillation frequency is therefore very small (or zero in an infinite
system), corresponding to a small (or zero) $V$ in
the benzene example. One way to think about
this feature in an extended system is by analogy
with picking up a rock, where we might suspect that
position-momentum uncertainty would prevent a solid object from being
localised \cite{anderson, anderson2}. The reason why the rock is stable is that we seek to
localise it within, say, a lattice spacing $a$, so the uncertainty in
momentum is of order $\hbar/a$ and the uncertainty in energy  for an
atom of mass $m$ is then $\Delta E = \hbar^{2}/2ma^{2}$. However, the key is that the solid
is made of $N$ atoms, {\it rigidly} stuck together, so $\Delta
E=\hbar^{2}/2Nma^{2}$ and, since $N$ is a macroscopic number, the energy
uncertainty is very small. This implies that the frequency of
oscillation between closely-spaced position states is vanishingly
small for a macroscopic piece of matter. The
analogous argument can be made in the case of an antiferromagnet (given
in the slightly more complicated terms of a
rigid rotor), and it turns out
that the closely-spaced energy levels in a macroscopic system
similarly allow a very stable wavepacket
to be constructed giving a stable
N\'{e}el AFM \cite{anderson2,matjaz}. The smoking-gun
experimental verification of its existence came from magnetic neutron
diffraction where the periodicity of the magnetic sublattices can be
directly seen from the measured Bragg peaks \cite{blundell}.

Although the AFM survives the strictures of quantum
mechanics, it's possible to postulate a state that shouldn't. 
1973 saw Philip Anderson's suggestion of a resonating valence bond
(RVB) state \cite{anderson3}. This is based on the notion of frustration, where it is
impossible to satisfy all of the interactions for a particular
geometry.\\
{\it Example}: consider a triangle decorated with Ising spins as shown in Fig.~\ref{fig:benz}(b) with
Hamiltonian $H = -\sum J_{ij} \sigma^{z}_{i}\sigma^{z}_{j}$,  where
all bonds favour antiferromagnetism ($J_{ij}=-J<0$). This state is frustrated as there is
no single state that satisfies all of the interactions. 
In this case the frustration gives
rise to six
lowest-energy spin configurations (with energy $E=-J$) and two excited states
(with $E=3J$). By contrast, an unfrustrated model where all $J_{ij}=J$ has a ferromagnetic, two-fold
degenerate ground state with energy $E=-3J$.
(The degeneracy here simply reflects the fact that all spins can align either
up or down.)
We see how frustration
increases not only the relative energy of the ground state but, more crucially, the
entropy through the large number of degenerate ground states that
result. The degeneracy becomes macroscopic in the thermodynamic limit
and has a destabilising effect on ordering.

To form Anderson's RVB state \cite{anderson3,savary,senthil,zhou}
we  extend the triangle to a spin-1/2 Heisenberg model on an infinite
two-dimensional triangular  lattice
with purely antiferromagnetic interactions on all bonds. We also upgrade
from an Ising spin model to a model of Heisenberg spins.\footnote{Reminder: two Heisenberg
  spins $|s_{1}s_{2}\rangle$ with isotropic antiferromagnetic exchange
  coupling between them will have a singlet ground state: $\left(|\uparrow\downarrow\rangle -
  |\downarrow\uparrow\rangle\right)/\sqrt{2}$, and a triplet of excited
states, comprising $|\uparrow\uparrow\rangle$,
$|\downarrow\downarrow\rangle$ and
$\left(|\uparrow\downarrow\rangle +
  |\downarrow\uparrow\rangle\right)/\sqrt{2}$.} 
Note first that, quantum mechanically, the ground state of a single
antiferromagnetic
bond is represented by an entangled spin
singlet $\left(|\uparrow\downarrow\rangle -
  |\downarrow\uparrow\rangle\right)/\sqrt{2}$. We therefore form a typical state
by decorating the lattice with singlets coupling nearest
neighbours [Fig~\ref{fig:benz}(c)]. We then repeat with a different choice of singlet
bonds to form another state with the same energy
[Fig~\ref{fig:benz}(d)]. We continue this process until we have all
possible coverings,  and then form our
final trial wavefunction by adding together all of
these states to form the RVB state. 
Notice how the RVB bears a resemblance to the benzene example, albeit
in an infinite system: in both we
can think of 
the choice of bonds resonating between different degenerate
configurations. 
Ultimately, the RVB state did not turn out to describe a $s=1/2$
Heisenberg antiferromagnet on a triangular lattice.\footnote{It has, however, since been invoked in several
contexts, not least to describe high $T_{\mathrm{c}}$ cuprate
superconductors.}
However, the RVB is our first example of a QSL. It does not break
rotational symmetry since each bond represents a $s=0$ spin
state. Moreover, the
state itself  can
be thought of in terms of a complicated dynamic dance of singlet
bonds that represents the underlying pattern of quantum entanglement. 

The notion of a dynamic pattern of quantum correlations that underlies a
magnetically-disordered state of macroscopic matter will now be
expanded via a consideration of some of the background concepts needed
to understand the QSL picture. 

\section{Ingredients of a QSL}\label{sec:ingredients}

There are a number of ingredients of QSLs that we describe here and
will feature in the rest of the article. 
The first  is {\bf reduced dimensionality}. When interactions are
confined to less than three spatial dimensions (3D), fluctuations become more
effective in destabilising order. In fact, the Mermin-Wagner
theorem \cite{blundell, lancaster}
says that for spins with a continuous degree of freedom, magnetic order will
always be destabilised above $T=0$. 
We shall mainly be concerned in this review with the case of two
spatial dimensions (2D) and time [known as (2+1) dimensions].
Although the two-dimensional Heisenberg magnet will not order for
$T>0$, the ferromagnetic certainly does at $T=0$, and there is good evidence
that the antiferromagnet should order at $T=0$ as well \cite{manousakis}. We therefore need some further
source of fluctuations if we want to prevent long-range magnetic order
and instead
promote the quantum order that gives robust (i.e.\ $T=0$) spin-liquid
behaviour. Frustration, as considered above for the triangle, is one possibility, but
there are others, including higher-order interactions or
disorder. (There are also QSL states that can be realised in 3D,
which we return to at the end of the article.)

The second ingredient is {\bf topology} \cite{lancaster,wen}.
Topology enters many-body physics in a number of ways. One is in
theories: some theories make no contact with the distances and time
intervals in a problem (i.e.\ the geometric features) and instead rely only on the underlying shape of
their space. These are topological theories: an example is the
Chern-Simons theory used to describe the FQH fluid, along with
some QSL states. A consequence of this lack of geometry turns out to
be that the Hamiltonian of a purely topological theory (such as a
the Chern-Simons Hamiltonian) is
$\hat{H}=0$, meaning all states are degenerate with zero
energy. The exact size of the large resulting degeneracy depends on the
topology of the underlying space (or manifold) on which we're working.
The topology of the manifold is an interesting case in itself.
If we impose periodic boundaries on our 2D lattice in both spatial
directions then the shape of the underlying topological space 
is a torus. This topology leads to a robust degeneracy:
it is very difficult to split the energies of the degenerate states with
perturbations.
Topological order (where there is a nonzero energy gap between the
ground state and all of  the excited states) follows
from the existence of these robust topological degenerate ground
states, and
is sometimes described in terms of  topology ``protecting'' the
gaps that appear in topologically-ordered states. 
A final appearance of topology is made by excitations \cite{lancaster,lancaster2}. Some
excitations are extended in space and cost a large amount of energy to
remove. 
 Topological excitations, often called topological defects,  include the domain wall in
the one-dimensional (1D) magnet, the vortex in 2D  and the monopole in
3D.
In each case, the large energetic penalty involved in
removing a topological excitation can be traced back to their shape (e.g. for the domain
wall, we would need to flip a semi-infinite number of spins to remove
the wall). 

{\bf Entanglement} is our next ingredient.  A state is entangled if it
is in a quantum superposition that cannot be written
as a product, even under an arbitrary (local) change of basis
states. The key example to have in mind here are two spins in an
entangled singlet state $\left(|\uparrow\downarrow\rangle- |\downarrow\uparrow\rangle\right)/\sqrt{2}$.
As we've said, quantum order describes the pattern of entanglements in
a many-body system, although it's rather hard to rigorously define a
many-particle entangled state. We saw one example in the RVB of
resonating singlets, we'll see another in our discussion of the toric
code model. 

The number of different {\bf elementary
  excitations}\footnote{The low-energy
  excited states that exist close to the ground state of an
  interacting, many-body system can be thought of as
hosting an assembly of elementary excitations. Examples of elementary
excitations are phonons in a crystal, magnons in a ferromagnet,
quasielectrons in a metal etc.} of the QSL are an important means of
classifying the states, especially whether they are {\bf gapped
  or gapless} (i.e. whether or not there is a nonzero energy gap
between the ground state and first excited state).
QSLs can support local particle-like excitations that can be fermionic,
bosonic or anyonic, and  can have  fractional quantum numbers.
For example, many QSLs support gapless, chargeless $s=1/2$
fermionic excitations known as {\bf spinons} \cite{giamarchi}.
In constrast, excitations can also be 
topological objects.
An example
of a (bosonic) excitation found in many QSLs is the topological {\bf vortex}.


In order to discuss spin liquids we shall also need some ideas from
quantum field theory (QFT) \cite{coleman,lancaster}. In a QFT we describe a system in terms of
a field (an object in which we input a position and output an amplitude
or quantum operator). Particle-like excitations in QFT are quantised excitations
in the field.
Operators in QFT act on  many-particle states that describe
the system. The most important state is the ground state or vacuum
$|0\rangle$. Excited states, such as particle excitations, are then added to this vacuum
state, using operators like $\hat{c}^{\dagger}$. For example,  to add  a
$c$-particle to the vacuum we write
$\hat{c}_{i}^{\dagger}|0\rangle = |1_{i}\rangle$, where $|1_{i}\rangle$ is a state
containing one $c$-particle at position $i$.
(We can also annihilate particles: $\hat{c}_{i}|1_{i}\rangle = |0\rangle$.)
One important model  to have in mind in our discussion is the tight-binding model,
describing the energies of electrons on a lattice with a Hamiltonian
\begin{equation}
  \hat{H} = \sum_{ij}(-t_{ij})\hat{c}^{\dagger}_{i}\hat{c}_{j},
  \label{eqn:tb}
\end{equation}
where the subscripts label lattice sites. 
The idea here is that the operators annihilate the electron at site $j$ and
create it at site $i$, causing an electron to hop between sites, making a contribution of kinetic energy of
$(-t_{ij})$. 
This model can often be solved in momentum space to give
$  \hat{H} =
\sum_{\boldsymbol{k}}E(\boldsymbol{k})\hat{c}^{\dagger}_{\boldsymbol{k}}\hat{c}_{\boldsymbol{k}}$,
where $E(\boldsymbol{k})$ is the electron dispersion. We'll
see that some important
models of QSLs can be boiled down to variations of this basic picture. 

The final, crucial, ingredient is {\bf gauge field theory} \cite{lancaster}.
A gauge field is one where there is some redundancy of
description. This is to say that the same physical
state can be described by several different configurations
of a gauge field. (You can think of this as a little like a language
in which a physically-equivalent scene can be described in English, French, Arabic,
Hindi, Japanese etc.)
The most familiar gauge field is the electromagnetic gauge field,
whose components are $A^{\mu} = (V, A^{x}, A^{y}, A^{z}) = (V,
\boldsymbol{A})$. This is often also called the electromagnetic potential.
We define the electric field $\boldsymbol{E}$ in terms of the
components of the  gauge field as
$\boldsymbol{E} = -\boldsymbol{\nabla} V - \frac{\partial \boldsymbol{A}}{\partial t}$, 
and the magnetic field $\boldsymbol{B}$ as
$\boldsymbol{B}= \boldsymbol{\nabla}\times\boldsymbol{A}.$ 
An important property of these equations is that there is some freedom
in how  the components $A_{\mu}$ are specified. In fact, if we make
the  {\it gauge transformations}
\begin{equation}
  \begin{array}{cc}
  V \rightarrow V - \frac{\partial\chi(t,\boldsymbol{x})}{\partial t}, &
  \boldsymbol{A}\rightarrow \boldsymbol{A} + \boldsymbol{\nabla}\chi
  \end{array}
  \end{equation}
where $\chi(t,\boldsymbol{x})$ is some arbitrary function, then the values for the
$\boldsymbol{E}$ and $\boldsymbol{B}$ fields are unchanged. 

The notions of anyons and of a gauge theory are so important that
we'll devote some more discussion to them in the following sections.

\section{Anyons and topology}\label{sec:anyons}

\begin{figure}
  \begin{center}
\includegraphics[width=4cm]{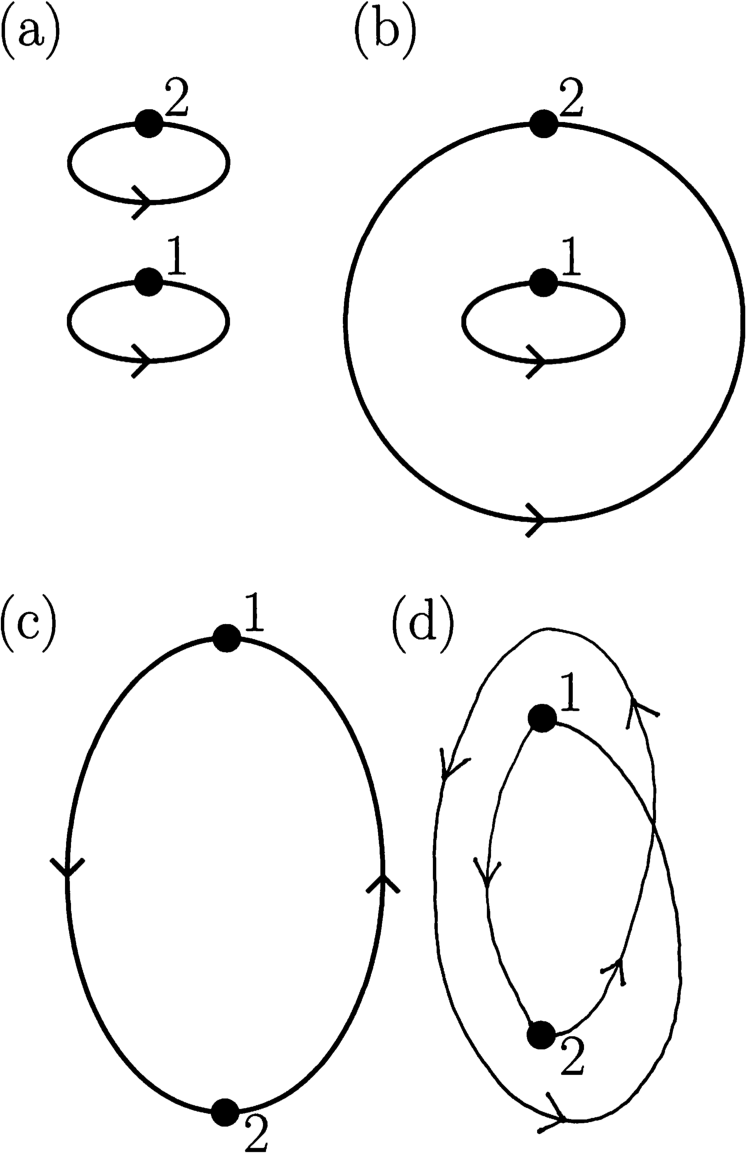}
\caption{
Examples of ways of exchanging particles. (a) and (b) are type-A
processes, where the particles end up at
the same positions. In case (b) one particle loops around the other
once in the exchanging process. (c) and (d) are type-B processes
where the particles exchange positions. In case (d) one particle loops
once around the other during the exchange. [Figure reproduced from
Ref.~\cite{lancaster}, reprinted with permission of Oxford University Press.]
\label{fig:anyons}}
\end{center}
\end{figure}

One of the most striking properties of QSLs is that they can host
excitations that are neither conventional fermions nor bosons \cite{lancaster}.
In quantum mechanics in 3D, changing the labels on two identical particles
results in a change of 
the wavefunction according to the rule
$\psi(x_{1},x_{2}) = \pm \psi(x_{2},x_{1})$,
where the $+$ sign applies to bosons and the $-$ sign to fermions. 
For the case of 2D space, the exchange of particles needs more careful attention. 
We start with two identical particles at positions $x_{1}$ and $x_{2}$
and identify two distinct ways of exchanging them.
The result of processes of type A is to move $x_{1}\rightarrow x_{1}$ and $x_{2}\rightarrow x_{2}$.
Some examples are shown in Fig.~\ref{fig:anyons}(a) and (b).  
The particles end up where they were originally, although they may move around each other. 
The result of type-B processes is to move $x_{1}\rightarrow x_{2}$
and $x_{2} \rightarrow x_{1}$ [Fig.~\ref{fig:anyons}(c) and (d)]. 
Again particles may move around each other several times before
settling at their final positions [as in Fig.~\ref{fig:anyons}(d)]. 

We ask what the relative quantum mechanical phase difference is between processes of type A and type B. 
The key parameter is the angle that one particle is moved around the other. 
Topology comes in here: given a set of particle paths, we can smoothly distort the paths of the particles, 
but we may not change the number of times particles wrap around each other without introducing 
singularities in the paths. 
Processes of type A involve rotating particle 2 around particle 1 by angle $\phi=2 \pi p$, 
where the winding number $p$ takes an integer value (including zero). Processes of type B involve rotations of $\phi = 
\pi (2p+1)$. Each value of $p$ describes a topologically distinct process. 
We suppose that these topologically distinct processes make
 a multiplicative contribution to the wavefunction of 
$\Phi(\phi)$ which is pure phase. 
If we carry out a sequential string of these processes then we require that the angles add, whilst the 
wavefunctions
should multiply. That is, $\Phi(\phi_{1} + \phi_{2}) = \Phi(\phi_{1})\Phi(\phi_{2})$, which
implies that $\Phi(\phi) = {\rm e}^{\mathrm{i} \eta \phi}$, where $\eta$ is a parameter which, crucially, doesn't need
to be an integer. 

Compare our upgraded 2D exchange to the old-fashioned definition. 
If we carry out the exchange
$(x_{1},x_{2})\rightarrow (x_{2},x_{1})$
 [shown in Fig.~\ref{fig:anyons}(c)]
then the formal
definition of exchange tells us that the wavefunction
should be identical (for bosons) or pick up a minus sign (fermions). However, the new version merely tells us that
$\phi = \pi$, resulting in a phase factor of $\Phi(\phi)={\rm e}^{\mathrm{i} 
  \eta \pi}$.
The two versions of exchange are only identical for the special cases that 
(a) we have $\eta = \mbox{even integer}$, when we recover the expected 
exchange behaviour for bosons or (b) we have $\eta =\mbox{odd integer}$, 
when we recover fermion exchange. 
However, this analysis shows that there are many more possible values of $\eta$
in 2D, since it doesn't have to be an integer.  We are therefore not tied
simply to bosons and fermions, the freedom to choose $\eta$ means we
can have any exchange statistics, and particles with such statistics
are called (Abelian) anyons.\footnote{Having only two spatial dimensions 
is vital to the argument. In 3D
all type-A processes are topologically identical since they are all deformable into paths where
the particles don't move. Similarly all type-B processes are topologically identical
and may be reduced to a simple exchange of particles. This reduction occurs 
because the extra dimension allows us to move the paths past each other in the third dimension,
shrinking all loops to zero. }

\section{Gauge theory}\label{sec:gauge}

QSLs are often named with a mathematical group (e.g.\ a $Z_{2}$ spin liquid, or
$U(1)$ spin liquid). This is related to the properties of the QSL's gauge
structure, considered in this section through a number of examples. 

{\it Example 1}: Consider complex scalar field theory, which is a
model built from continuous complex-number valued fields $\psi(x)$ that are
described by a Hamiltonian
\begin{equation}
\hat{H} = (\partial_{x}\psi)^{\dagger}(\partial_{x} \psi) +
m^{2}\psi^{\dagger}\psi, 
\end{equation}
where $m$ is a constant.
This theory has an internal symmetry known as a {\it global phase symmetry}, or global $U(1)$
symmetry. This is a shorthand for the observation that with the replacement
$\psi(x) \rightarrow \psi(x){\rm e}^{\mathrm{i}\alpha}$,
the Hamiltonian $H$ does not change.\footnote{This transformation generates
  the elements of the Lie group $U(1)$, hence the name.}
This is a global transformation in that the phase changes by the same
amount ($\alpha$) at every point.

What if we attempt to change the
phase {\it locally}, i.e.\ differently at each point?
That is, change the phase by an amount
$\alpha(x)$ that depends in some arbitrary manner on position.
Under the local transformation, the derivatives change as follows:
\begin{align}
\partial_{x}\psi(x) \rightarrow
\partial_{x}\left[\psi(x){\rm e}^{\mathrm{i} \alpha(x)}\right]
=& 
{\rm e}^{\mathrm{i} \alpha(x)}  \partial_{x}\psi(x)
+ \psi(x){\rm e}^{\mathrm{i} \alpha(x)} \mathrm{i}\partial_{x}\alpha(x) \nonumber\\
=& {\rm e}^{\mathrm{i} \alpha(x)}
\left\{\partial_{x} +\mathrm{i} \left[\partial_{x}\alpha(x)\right]\right\}\psi(x).
\end{align}
Similarly
$\partial_{x}\psi^{\dagger}(x) \rightarrow
{\rm e}^{-\mathrm{i} \alpha(x)}\left\{\partial_{x} 
-\mathrm{i}
\left[\partial_{x}\alpha(x)\right]\right\}\psi^{\dagger}(x)$, and
the first term in the Hamiltonian becomes
\begin{equation}
(\partial_{x}\psi)^{\dagger} (\partial_{x}\psi)
-\mathrm{i}\psi^{\dagger} (\partial_{x}\alpha) (\partial_{x}\psi)
+\mathrm{i}\psi  (\partial_{x}\psi^{\dagger}) (\partial_{x}\alpha)
+\psi^{\dagger}\psi (\partial_{x}\alpha) (\partial_{x}\alpha),
\end{equation}
which is not what we started with.
Perhaps unsurprisingly, the theory is not therefore 
invariant with respect to local phase transformations. However, to fix
things
we introduce a new {\it gauge field} whose job is to cancel out the effect of the change in
internal variable $\alpha(x)$ with position.
This enters into the Hamiltonian as a covariant field
derivative $D_{x}$, defined by
$D_{x}\psi(x) = \partial_{x}\psi(x) - \mathrm{i}qA^{x}(x)\psi(x)$,
where $q$ is a coupling constant. 
We define the gauge field such that if we change the phase by angle
$\alpha(x)$,  the components of the gauge field transform according to 
$A^{x}(x) \rightarrow A^{x}(x) + \frac{1}{q}\partial_{x}\alpha(x)$. 
Now if $\psi(x)\to\psi(x) {\rm e}^{{\rm i}\alpha(x)}$, then
\begin{align}
D_{x}\psi = (\partial_{x}- \mathrm{i}q A^{x})\psi 
 \to 
{\rm e}^{\mathrm{i}\alpha}(\partial_{x}\psi) + \mathrm{i}\psi {\rm e}^{{\rm i}\alpha}(\partial_{x}\alpha)
-\mathrm{i}qA^{x}\psi {\rm e}^{\mathrm{i}\alpha} - \mathrm{i}\psi {\rm e}^{{\rm i}\alpha}(\partial_{x}\alpha)
= {\rm e}^{\mathrm{i}\alpha} D_{x}\psi.
\end{align}
This property makes the whole Hamiltonian invariant under local phase changes if we replace ordinary derivatives
by covariant ones:
$\hat{H} = (D^{x}\psi)^{\dagger}(D_{x}\psi) + m^{2}\psi^{\dagger} \psi,$
since now with $D_{x}\psi \rightarrow {\rm e}^{\mathrm{i} \alpha} D_{x}\psi $, the first
term is invariant. 

The message of these manipulations is that if the phase $\alpha(x)$ is a
function of $x$ then, in order to guarantee local phase invariance, a new
gauge field $A^{x}(x)$ is required to form the covariant
field derivative. Furthermore, we recognise this field as akin to the
electromagnetic gauge field,
where we saw that one of its features was that,
without changing $\boldsymbol{E}$ and $\boldsymbol{B}$, the gauge field
could be changed by an arbitrary amount
$A^{x}(x) \rightarrow A^{x}(x) + \partial_{x}\chi(x)$. 
If we identify $\chi(x)$ with
$\alpha(x)/q$, then the transformation demanded by our argument
is simply electromagnetic gauge invariance, confirming that $A^{x}$ has the
usual properties of a gauge field as defined in electromagnetism.
We say that the gauge field has its own dynamics, which in
electromagnetism are governed by the Maxwell equations, written in
terms of the (3+1)-dimensional field $A^{\mu}$. 


{\it Example 2:} In Section~\ref{sec:spinons} we will introduce a
model of spinon particles
on a lattice
represented in terms of the operator $\hat{f}^{\dagger}_{i}$, which
creates a spinon at site $i$.
This is similar to the continuous model in Example 1, except
that position is now a discrete variable.
We again demand invariance under a
local $U(1)$ phase transformation $\hat{f}_{i}\rightarrow\hat{f}_{i}{\rm
  e}^{\mathrm{i}\phi_{i}}$, which changes the phase at each lattice
point $i$. To ensure this we again introduce a
gauge field, this time $\bar{\chi}_{ij}{\rm e}^{\mathrm{i}\theta_{ij}}$, where $\theta_{ij}$
is a phase that depends on the bond between sites $i$ and $j$.
The spinons and the gauge field interact according to a 
Hamiltonian $\hat{H} = \hat{f}^{\dagger}_{i}\hat{f}_{j}\left(\bar{\chi}_{ji}{\rm
  e}^{\mathrm{i}\theta_{ij}}\right)$, so 
if
the gauge field
transforms according to
$\theta_{ij}\rightarrow\theta_{ij}+\phi_{i}-\phi_{j}$, then 
the Hamiltonian is symmetric under local $U(1)$
transformations of the spinon operators. We will see later that this
gives us a $U(1)$ gauge theory of
spinons. 
  
{\it Example 3:} $Z_{2}$ (more often written $\mathbb{Z}_{2}$ in the
mathematics literature) is a group
  with two elements: $1$ and $-1$. For the Kitaev model on a lattice (Section~\ref{sec:kitaev}), we
  identify an operator $\hat{c}_{i}$ and demand invariance under a
  local transformation $\hat{c}_{i} \rightarrow W_{i}\hat{c}_{i}$,
  where $W_{i}$ is an arbitrary function that outputs 1 or $-1$ at
  the $i$th site. We identify a gauge field $\hat{u}_{ij}$ that depends on
  two positions $i$ and $j$, 
and a Hamiltonian
$\hat{H} = (\hat{u}_{jk})\hat{c}_{j}\hat{c}_{k}$. If 
$\hat{u}_{ij}$ transforms according to
  $\hat{u}_{ij}\rightarrow W_{i}\hat{u}_{ij}W^{-1}_{j}$, then this
  allows the theory to be locally $Z_{2}$ invariant.
  
  {\it Example 4:} Our discussion in the next section relies on another $Z_{2}$
  gauge field $S_{ij}$, 
  where positions are specified by two lattice sites $i$ and $j$.
    Here we will again transform via the local transformation $S_{ij}\rightarrow
  W_{i}S_{ij} W^{-1}_{j}$ in the same way as Example 3. This
  transformation leaves the Hamiltonian invariant, as we shall now discuss. 

\section{Our second spin liquid: the toric code}\label{sec:toric}
There is a class of models that allows us to get a handle on some of the
ingredients we've met so far: we start with the {\bf toric code}
and then (Section~\ref{sec:toric2}) consider a simplified version of this model. We shall see
how these models give rise to spin-liquid ground states with
topological ground-state degeneracy, anyonic excitations
and a gauge-field structure based on the group $Z_{2}$.

\begin{figure}
  \begin{center}
    \epsfig{file=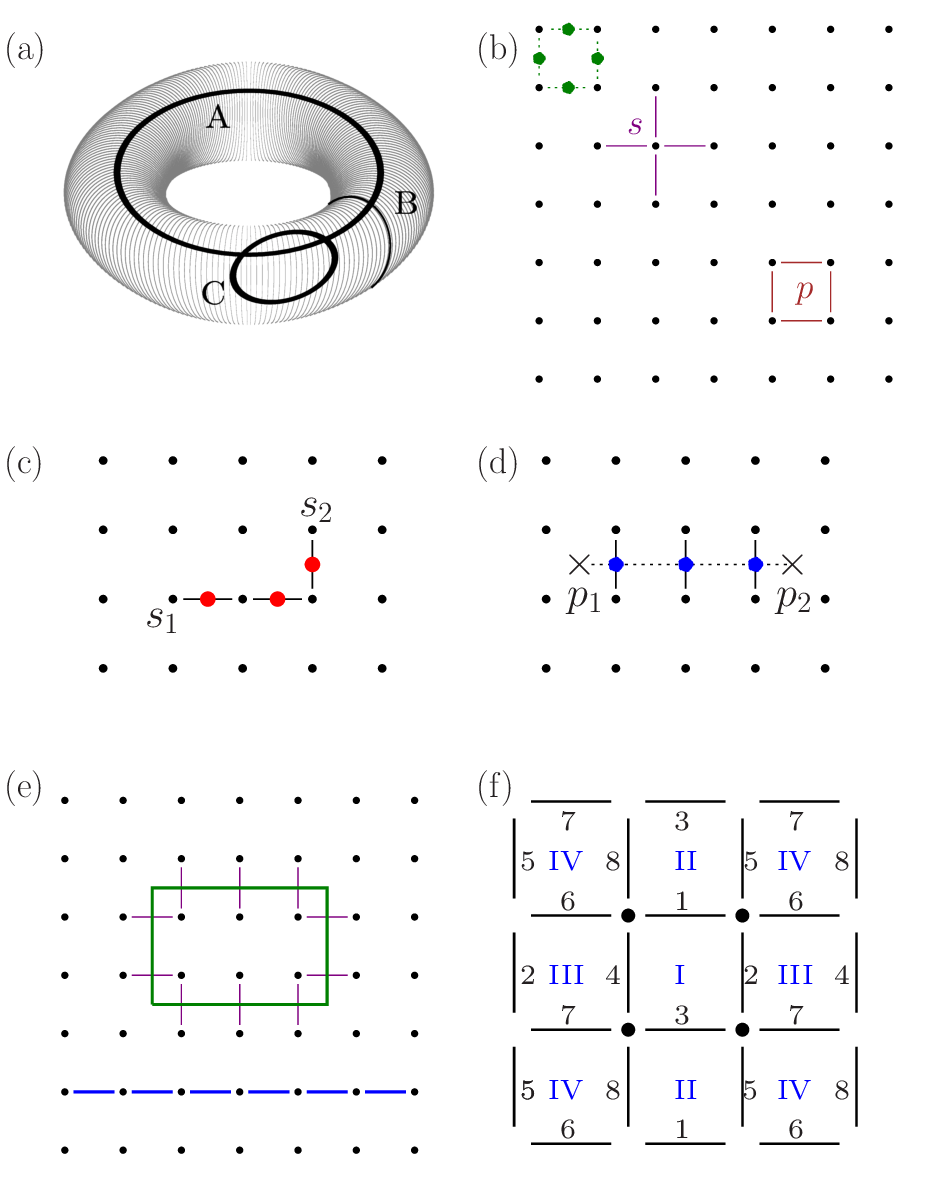,width=12cm}\\
    \caption{(a)
      A torus with three different sorts of paths on its
      surface. [Figure reproduced from Ref.~\cite{lancaster}, reprinted with permission of Oxford University Press.]
            (b) Square lattice in which spins lie in the spaces
      between lattice points. Example spin positions are shown in
      green in the upper left. 
An example plaquette $p$ and star $s$ are also
shown.
 (c) Flipping a connected
string of spins aligned along $\pm x$ (red dots)
excites two $e$ anyon (or star) excitations at the ends of the string.
(d) Flipping spins aligned along $\pm z$ along a string perpendicular
to the bonds (blue dots) creates two $m$ anyon (or plaquette) excitations.
(e) Top: the action of the star operator on a path through the lattice flips spins (purple), giving a loop of flipped
spins on the {\it dual} lattice (green). Bottom: a line of flipped spins (blue)
makes an $A$-type path on the torus.
(f) Configuration of bonds and plaquettes that results from applying
periodic boundary conditions to a lattice of 4 points. Arabic numerals
label bonds, roman numerals label plaquettes.  \label{fig:toric}}
\end{center}
\end{figure}

The toric code \cite{kitaev,savary} describes spins on a 2D
lattice with periodic boundary conditions. The boundary conditions
are a key feature: by identifying each of the two perpendicular
directions on the lattice, we obtain a space with the topology of a
torus (hence the name of the model). An example torus is shown in
Fig.~\ref{fig:toric}(a). This surface has two holes: one through the empty
middle of the doughnut, and one through the tubular part (filled with
dough in the edible version). The important thing to note is that there are
three main sorts of paths on the torus [see Fig.~\ref{fig:toric}(a)]: $A$-type and $B$-type paths wrap
the two different holes that characterise the torus, while $C$-type
paths wrap neither of the holes. It is impossible to smoothly deform
one type of path into a different one (that is, they are topologically
distinct: we would need to cut
the paths and glue them back together to do this). 

The lattice in the toric code model is decorated by
$s=1/2$ spins which are, rather unusually, situated on the
centres of the {\it bonds}, rather than on the vertices
of the lattice, as one would usually expect,
[the spins are represented by green dots in Fig.~\ref{fig:toric}(b)]. We shall mostly  work in a basis of
spins directed up and down along $z$, and operate on a spin between
sites $i$ and $j$
with
Pauli sigma operators using the usual rules $\hat{\sigma}_{ij}^{z}|\uparrow_{ij}\rangle=|\uparrow_{ij}\rangle$,
$\hat{\sigma}_{ij}^{z}|\downarrow_{ij}\rangle=-|\downarrow_{ij}\rangle$, $\hat{\sigma}_{ij}^{x}|\uparrow_{ij}\rangle=|\downarrow_{ij}\rangle$
and
$\hat{\sigma}_{ij}^{x}|\downarrow_{ij}\rangle=|\uparrow_{ij}\rangle$.

There are two contributions to the Hamiltonian [see Fig.~\ref{fig:toric}(b)]: (i) from each
square plaquette (labelled with an index $p$) on the lattice, and (ii) 
from each star, which is the cross-shaped structure formed from the nearest-neighbour bonds of
 lattice site $s$. The contribution from a plaquette
is given by
$\hat{P}_{p} = \prod_{ij \in p}\hat{\sigma}_{ij}^{z}$, 
which is to say a product of the $z$-components of spins on the bonds around a
plaquette. This takes values $\pm 1$. The contribution from a star is
given by
$\hat{R}_{s} = \prod_{ij \in s}\hat{\sigma}_{ij}^{x}$,
(i.e.\ operate with the $x$-sigma matrix, flipping spins in our
$\sigma^{z}$ basis).
This contribution also takes values $\pm 1$ (most easily seen by
transforming to a basis of spins directed along $\pm x$). 
Each of these two operators commutes amongst themselves and, as can be checked, they commute with each other. 
We put the two contributions together to form the
Hamiltonian for the toric code
\begin{equation}
  \hat{H} = -g\sum_{p}\hat{P}_{p} -t\sum_{s}\hat{R}_{s},
  \label{eq:toric_hamiltonian}
\end{equation}
where $g$ and $t$ are positive constants.

\subsection{Ground state and excitations}

On one level, the toric code is quite a simple model whose properties can be
guessed. 
We can immediately spot a candidate ground state $|0\rangle$: if we can find a
wavefunction such that all eigenvalues
$R_{s}=P_{p}=1$ we obtain the lowest-possible energy. Consider
plaquette $q$, to get $P_{q}=1$ we must have an even number of down
spins. (There are, of course, several ways to arrange this.)
We can also see the role of entanglement. Consider the ground state
expressed in a basis of eigenstates of the $\hat{\sigma}^{x}$
operator. Each spin is shared between two stars.
To make sure a star has $R_{s}=1$ there must be an even
number of  down (along-$x$) spins 
in the star. Moreover, in order that {\it all} stars have
$R_{s}=1$, these down spins must be 
must be arranged in closed loops. (Try it and see!)
The ground state $|0\rangle$ is an equal amplitude 
superposition of such loop states. 
Such an entangled ground state does not have a preferred spin direction and so is
magnetically disordered and is an example of a quantum spin liquid.

Excitations from the ground state have some plaquettes or stars
contributing $-1$ to the Hamiltonian. 
Excitations with $P_{p}=-1$ are known as vortices. 
These are
bosons, they cost energy $2g$ to produce, and are 
also known as magnetic ($m$) particles (or sometimes as visons). The other sort of
excitation is a single negative star $R_{s}=-1$, which is also a
boson,  this time  
costing energy $2t$, and is known as an electric ($e$) charge.
Neither $m$ nor $e$ can be created locally as
individual excitations using a single operator, as we do with
particles (where $c^{\dagger}|0\rangle = |1\rangle$). 
For the $e$ particles, for example, the best we can do is
flip a string of spins, which creates {\it two} localised $e$ particles: one at
each lattice site at the end of the string of flipped spins
[Fig.~\ref{fig:toric}(c)]. This is because each spin is shared between
two lattice sites, and hence shared by two stars.
The $m$ particles are also created in pairs, occurring at plaquette centres at the
ends of strings of flipped bonds, this time  as shown in
Fig.~\ref{fig:toric}(d). Again this is due to each 
spin being shared between two plaquettes. 
This non-locality of excitations leads to anyon statistics, although
in slightly more subtle form to what we've seen so far. 
Although $e$ and $m$ are both bosonic,
moving an $e$ around an $m$
gives a $\pi$ phase shift, resulting in a negative sign. We say that the particles have {\it mutual
  statistics}, specifically we call them mutual {\it semions} in this context.
Additionally, the composite particle $\varepsilon$ made
up of a boson $e$
and a boson $m$ is actually a fermion.



Using these ideas, we can deduce some basic properties of the
toric-code ground
state more formally. 
Working in a basis of eigenstates of $\sigma^{z}$ operators,
 we specify spin
configurations by writing a wavefunction $|\{S_{ij}\}\rangle$,
where $\{S_{ij}\}$ are a list of the eigenstates of each
operator $\sigma^{z}_{ij}$  for the bond between sites $i$ and $j$. The ground state
$|0\rangle$ will be built from a superposition of these  spin states $|\{S_{ij}\}\rangle$. 
To find $|0\rangle$, 
require for the plaquette term that
$\hat{P}_{p}|0\rangle=|0\rangle$. This is only possible for a state of
the form
\begin{equation}
  |0\rangle = \sum c_{s} \left|\{S_{ij}\}\right\rangle,
  \label{eq:toric_groundstates}
\end{equation}
where the $c_{s}$s are constants and the sum is constrained to be over those spin configurations that
feature no vortices (i.e.\ where $P_{p}=1$).

In this $\sigma^{z}$ basis, the star operators act on an $|\{S_{ij}\}\rangle$ to
flip spins. If we start with a state with no vortices and act with the
star operator on sites that lie on paths around the lattice
then we create a pattern of down
spins on an up-spin background. (The star operator never creates
vortices owing to the commutation of the two operations.) If we
visualise this by linking the down spins via the {\it dual lattice} (i.e.\ the
lattice formed from points in the middle of each lattice point on the
original lattice), the result is always closed
loops of down spins, of the sort shown in green in the example in Fig.~\ref{fig:toric}(e). A ground state
is formed from the superposition of all such vortex-free states comprising closed
loops of down spins on the dual lattice. These all contribute with
equal coefficients $c_{s}=c$ in eqn~\ref{eq:toric_groundstates}. 
The resulting state gives $\hat{R}_{s}|0\rangle=|0\rangle$, confirming
that we 
have a ground state.

Although this accounts for one ground state, there are actually three
more degenerate states, making
four in total. 
To access the others, let's consider
the {\it Wilson loop operator} $\hat{U}(\mathcal{C})$, which
multiplies the spins around a closed contour $\mathcal{C}$. The operator is
given explicitly by
\begin{equation}
\hat{U}(\mathcal{C}) =
\hat{S}_{ij}\hat{S}_{jk}...\hat{S}_{li},
\label{eq:wilson_loop}
\end{equation}
where the indices are selected to take us around the contour. 
The eigenvalue $U(\mathcal{C})$, which can take values $\pm 1$, is called the $Z_{2}$ flux
through $\mathcal{C}$. (For example, the flux from a vortex excitation
(i.e.\ $P_{p}=-1$ on a
plaquette $p$) is $U=-1$.)
Since the star operator flips {\it pairs} of spins along any contour through
the lattice, it must
commute with $\hat{U}(\mathcal{C})$. Now, if we take the path around the two distinct
paths that wrap a torus [$\ell_{1} = A$ or $\ell_{2}=B$ in Fig.~\ref{fig:toric}(a)], then we must have
$U(\ell_{i})=\pm 1$.
[The ground-state configuration we formed in
the last paragraph
has $U(A) = U(B)=1$.]

The four degenerate ground states come from the possibility of adding
loops of down spins that wrap all the way around the torus in either of those two distinct
paths that link the torus [i.e.\ paths $A$ or $B$ in Fig.~\ref{fig:toric}(c)]. An
example of an $A$-type loop of down spins is shown in
Fig.~\ref{fig:toric}(e). A $B$-type contour that wraps the
torus must cross this line of down spins and yield $U(B)=-1$ [we also have
$U(A)=+1$]. Note that there is no way to create the
down-spin loop from the application of the star operators,
and also that the state containing just
this loop has no vortices.  Compared to the state with no down-spin
loops that wrap the torus, this state will give distinct patterns  of down spins
on the dual lattice once we
start working on it with the star operator. As a result, the state
formed by all down-spin loops in an $U(A)=1, U(B)=-1$, system yields a second
degenerate ground state. The other degenerate states have $U(A)=-1, U(B)=+1$
[i.e.\ a vertical line of flipped spins on Fig.~\ref{fig:toric}(e)]
and $U(A)=U(B)=-1$ [i.e. both a vertical and horizontal line of flipped
spins].

The resulting 4-fold degeneracy is a {\it topological}
degeneracy, meaning that it occurs for all lattices on the torus.
In the language of flux, we might call $U(A)$ the flux through
one hole of the torus and $U(B)$ the flux through the other. States
with $U=1$ have no flux through the torus while states with $U=-1$
have one unit of flux threading it.\footnote{In fact, for a genus $g$
  Riemann surface, we can put zero or one unit of flux through each of
the 2g holes of the surface, leading to a $2^{2g}$ degenerate ground state.}
The degeneracy is very robust, as we can see if we treat the star term
as a small perturbation to the vortex term by assuming $g\gg t$ in eqn~\ref{eq:toric_hamiltonian}. 
Recall from degenerate perturbation theory that to compute the
splitting caused by a perturbation,  we need to identify a
series of operators that takes the ground state up to an excited state
(or states) and then returns it back to another state in the ground-state manifold. 
 The only way this can occur here using the star operator is for
it to create excited stated by flipping spins all of the way around the torus.  For an $L\times L$
 lattice, this requires $L$
excitations. Perturbation theory tells us that any energy gap caused by
such a perturbation is
of order $\Delta E\approx t^{L}/g^{L-1}$, but as we tend to the
thermodynamic limit of a system
$L\rightarrow\infty$, we find $\Delta E\rightarrow 0$. We conclude
that the topological degeneracy is robust. In fact, the degeneracies
in our toric code are exact, and not split at all by $ t^L/g^{L-1}$ terms, although
such a splitting would apply in a more generic model.

\subsection{Counting states in $Z_{2}$ theory}\label{sec:toric2}
{\it This subsection, and the following one, are a tutorial given in
  terms of a simplified
  model theory to explicitly compute the degeneracy via a
  demonstration of
  the gauge structure. You can skip to
  Section~\ref{sec:spinons} at this stage if desired.}

Let's simplify the toric code model further and consider only the
plaquette term \cite{wen} 
 with spins acted on by operators
$\hat{S}_{ij}=\hat{\sigma_{ij}^{z}}$ (that is, we set $t=0$ in
eqn~\ref{eq:toric_hamiltonian}). 
The Hamiltonian is now simply $\hat{H} = -g\sum_{p}\hat{P}_{p}$,
where $\hat{P}_{p} = \prod_{ij\in
  p}\hat{S}_{ij}$
now operates on the spins in a plaquette $p$ only. 
For definiteness take the number of sites on the
lattice to be $n=4$ so we have only a single
plaquette $p=\mathrm{I}$.
In the absence of boundary conditions, the spins on this plaquette can adopt
$2^{n} = 16$ configurations, half giving $P_{\mathrm{I}}=1$; half
giving $P_{\mathrm{I}}=-1$.

Now apply the all-important boundary conditions
and get Fig~\ref{fig:toric}(f) with $2n=8$ spins on the resulting
bonds (labelled 1-8 in the figure), and
therefore $2^{2n}=256$ possible spin configurations. However, the
description in the Hamiltonian $\hat{H} = -g\sum_{p}\hat{P}_{p}$ is not
given at the level of spins, but at the higher level of plaquettes.
By seeking to describe the system's states at this higher
level, 
we'll see that a description in terms of $P_{p}$ 
can be visualised as a pattern of fluxes that thread the lattice. 
 Compared to the model without boundary conditions, the periodic boundaries 
 provide three additional plaquettes, making four in total [labelled
 I, II, III and IV in Fig.~\ref{fig:toric}(f)]. However,
 these four plaquettes are not independent, since each spin is shared between two
 plaquettes. If we multiply all of the plaquette operators together,
 each spin features twice in the product, and since $P_{p}=\pm 1$, we have
the constraint $\prod_{p}P_{p}=1$. A consequence is that there are only $3$
possible values of energy $E=\sum_{p=\mathrm{I}}^{\mathrm{IV}}P_{p}$,
namely $4, 0 $ and$ -4$.
There is one set of $\{P_{p}\} = (P_{\mathrm{I}},P_{\mathrm{II}},P_{\mathrm{III}},P_{\mathrm{IV}})$ that leads to $E=4$,
one set that gives $E=-4$, and six sets with $E=0$, making 8 different
allowed sets (called ``patterns of flux'' below)
in total. This is a lot less than the 256 possible spins states!
In fact, consulting Table~\ref{table}, we find that there are $8\times
2 \times 2=32$ ways
of arranging spins to get each allowed choice of $\{P_{p}\}$.\\
{\it Takeaway: there are $2^{2n}=256$ configurations of spins in the model. There
  are only 8 unique sets of $\{P_{p}\}$. That is, there are 8 distinct
  patterns of flux through the lattice. It's therefore impossible 
  to uniquely label spin configurations with a list of $P_{p}$. }

\begin{table}
  \begin{center}
  \begin{tabular}{|ccc|}
    \hline
    Plaquette & Spins in  & No.\ of free choices to \\
    $p$ & the plaquette & obtain a given $P_{p}$\\
    \hline
    I & 1 2 3 4 & 8\\
    II &  1 3 5 8 & 2\\
    III & 2 4 6 7 & 2\\
    IV & 5 6 7 8 & 0\\
                   \hline
  \end{tabular}
  \caption{Decorate the periodic lattice in Fig.~\ref{fig:toric} with
    spins to realise a pattern of $(P_{\mathrm{I}}, P_{\mathrm{II}},
    P_{\mathrm{III}}, P_{\mathrm{IV}}$), in the order $p=$I, II, III,
    IV. There are 8 choices to obtain $P_{\mathrm{I}}$, but because
    spins are shared, these are constrained for the other plaquettes.
    The table shows that there are $8\times2\times2=32$ spin
  configurations that give each  pattern of fluxes $\{ P_{p} \}$.  (We check that $32\times
8$ different $\{P_{p}\}$ gives our 256 possible spin configurations.) \label{table}}
\end{center}
\end{table}

Although there are 8 patterns of flux, we don't actually know how many
states exist in the Hilbert space of the system when it is described in
terms of fluxes.
This is because both degeneracy and gauge equivalence will cause there to be several
spin states that give an equivalent pattern of fluxes.

Evaluating the balance between degeneracy and gauge equivalence
is the subject of the next subsection. However, if formal gauge
theories are not to your taste, here is the direct argument: our
simplified model has $t=0$ in eqn~\ref{eq:toric_hamiltonian}. We might ask how the
energies of these sets of states are split for $0 < t/g \ll 1$. The
answer is that all the exact eigenstates for $t, g > 0$ can be
labelled by the positions of the $e$ and $m$ particles, plus the values of
the fluxes around the two loops that encircle the torus. Here we have
8 configurations of $m$ particles  (i.e.\ one with no plaquettes occupied, one with all four plaquettes occupied, and six with two of the plaquettes occupied)
, and for each set of $m$ particles
there are 8 possible configurations of $e$ particles (i.e.\ one with no
sites occupied, one
with all four sites occupied, and six with two of the sites
occupied). For each of these 64 configuration of $m$ and $e$ particles
we've seen that there are
4 possible configurations of flux that  wrap the torus.
So the total number of states is $8 \times 8 \times 4
= 256$ as expected. For $t, g > 0$ these states must exist as 4-fold degenerate levels,
since the values of the flux don't influence the energy of the states.

\subsection{Introducing gauges}

Many of the 256 spin states are gauge
equivalent.
Recall that ``gauge equivalence'' is a statement that
several different configurations specified by a gauge field can
correspond to the same state in the Hilbert space of a
system.
We see this if we make a local
transformation: the transformed system gives the same Hamiltonian and
observables.
For our problem, the Hamiltonian is given in terms of the configuration of plaquettes,
which therefore determines the Hilbert space. We will call
these states in the Hilbert space, which can be visualised in terms of
the patterns of fluxes through the plaquettes, the {\it physical
  states} for this problem. 
Since many of the  underlying spin states correspond to the same
pattern of fluxes, we can therefore treat the specific description of spins as
a gauge field.
To summarise, on putting
the spins on bonds and specifying the configurations using the
the plaquette operators $\hat{P}_{p}$ in our Hamiltonian, rather than
the spins, 
we obtain a redundancy of description, and hence we say that a gauge structure
{\it emerges}.\footnote{The situation here is subtly different to that
of electromagnetism, where there is no direct experimental access to
the gauge field. Here we could imagine making measurements of the
spins in a real system via magnetic susceptibility. However, by treating the physics at
the (Hamiltonian) level of plaquettes and, equivalently, fluxes, there's a sense in which we're
treating the spins as a microscopic description that we can't access, and working
instead with what
emerges at the more coarse-grained level of fluxes. }


In our model, $S_{ij}$ is the gauge
field and we'll see that several states described by different
$S_{ij}$
correspond the same 
state in the Hilbert space of the model when described in terms of
plaquette operators. 
 Technically, we can group spin states into collections called  {\it gauge-equivalent
classes}, where  all of the states in such a class are related by a local
transformation and so represent the same physical state. As a result,
states in a Hilbert space are in a 1-1 relation with
the number of gauge-equivalent classes.

The gauge group for our model is $Z_{2}$, which is to say specifically that a local
transformation $W_{i}$ of the spin field $S_{ij}$  leaves the Hamiltonian
invariant. Here the local transformation $W_{i}$ is an arbitrary
function that takes
values $\pm 1$ at the
 different lattice sites. 
Two spin fields $S_{ij}$ and $\tilde{S}_{ij}$ are {\it gauge equivalent} if they
are related by a local transformation $W_{i}$ via
$\tilde{S}_{ij} = W_{i} S_{ij} W^{-1}_{j}$.
You can see an example of how this works by selecting some site $k$ and setting
$W_{k}=-1$ and all of the other $W_{i}$ to be $+1$. This flips spins
in a star shape around site $k$, but since two of the flipped bonds
feature in each plaquette surrounding $k$, the Hamiltonian $\hat{H} =
-g\sum_{p}\hat{P}_{p}$
is gauge invariant: you get the same value of energy, no matter which
$S_{ij}$ you pick.
Moreover, you can make $2^{n}=16$ different $W_{i}$ functions on our
lattice, so this is the number of possible local transformations. 

In order to work out if two states on a lattice are really the same physical state
(i.e.\ if they are gauge equivalent),
an important role is played  by the Wilson-loop operator
$\hat{U}(\mathcal{C})$ from eqn~\ref{eq:wilson_loop}.
 The key is that $U(\mathcal{C})$ itself is gauge invariant, so states that give different values of $U(\mathcal{C})$ cannot be
related by an arbitrary local transformation. Therefore if we know the pattern
of fluxes provided by evaluating $U(\mathcal{C})$ around each plaquette,
then we have the physical description of a state.
Note that if $\mathcal{C}$ is a single plaquette $p$ then $\hat{U}(\mathcal{C})\equiv
\hat{P}_{p}$, and so the sets
$\{P_{p}\}$ pick up the flux through each plaquette and label states
in a way that's invariant with respect to local transformations. In
short: a
different $\{P_{p}\}$ means a different physical state.
However, the same $\{P_{p}\}$ doesn't guarantee that all of the states
are gauge equivalent (i.e. the same physical state) because we don't yet know how many states are distinct
states that are degenerate.
Returning to our example, we have that a number (32 in our case) of spins states will
give the same pattern of fluxes described
by a given $\{P_{p}\}$.

To finally compute the degeneracy, we need a technical argument \cite{wen}, which goes as follows:
{\it There are $2^{n}=16$ different possible
local transformations, but
there are two special transformations that don't change the spin configuration,
namely  $W_{i}=1$ and $W_{i}=-1$, for all $i$.
A (non-trivial) consequence of the
existence of these two special transformations is that the 16 gauge
transformations can actually only create $2^{n}/2=8$  gauge-equivalent
configurations, where the 2 here is the number of these special
transformations. }
With this key fact in hand, we can say that there are $2^{2n}/(2^{n}/2)=2\times 2^{n}=32$ gauge-equivalent classes [i.e.\ total  number of spin states (256), divided by number
of gauge-equivalent configurations (8)] and therefore 32 different
physical states in the Hilbert space. So out of 256 total spin
configurations, only 32 are distinct physical states. 
To label the flux states we use the $Z_{2}$ flux
through a plaquette provided by the eigenstates of the $\hat{P}_{p}$
operator. We saw
that there are
only $2^{n}/2=8$ different values of $\left\{P_{p}\right\}$ and so
only 8 distinct patterns of flux through the system. 
However, if each state is 4-fold degenerate we are saved, since
then the different flux patterns give the $8\times 4=32$
physical states in the Hilbert space.

\noindent
{\it Takeaway: There are 32 physical states in the Hilbert space of the model. Each
  of these states corresponds to 8 gauge-equivalent states (accounting
  for 256 spin configurations).
There are
  8 different patterns
  of flux encoded by $\{P_{p}\}$. Each of the 8 flux patterns corresponds to  4
  physically-distinct degenerate states.\footnote{Since each of the physically distinct states can
  be expressed as 8 gauge equivalent configurations of spins, 
  there are 32 spin configurations giving
  each pattern of fluxes, as we found above.}}

Let's pause and consider what we've learnt from this technical discussion. The
simplified toric-code model has a spin-liquid ground state which
respects rotational spin symmetry and is highly entangled. The model
is invariant with respect to local $Z_{2}$ transformations, giving it
a rich gauge structure. The topology of the lattice, via the gauge structure, causes all states
to be 4-fold degenerate. Finally, we might congratulate ourselves by
getting this far with a quotation from X-G.~Wen \cite{wen}:
\begin{quotation}
If you feel the definition of $Z_{2}$ gauge theory is formal and the
resulting theory is strange, then you get the point.
\end{quotation}

\section{Spinons}\label{sec:spinons}
\begin{figure}
  \begin{center}
    \includegraphics[width=12cm]{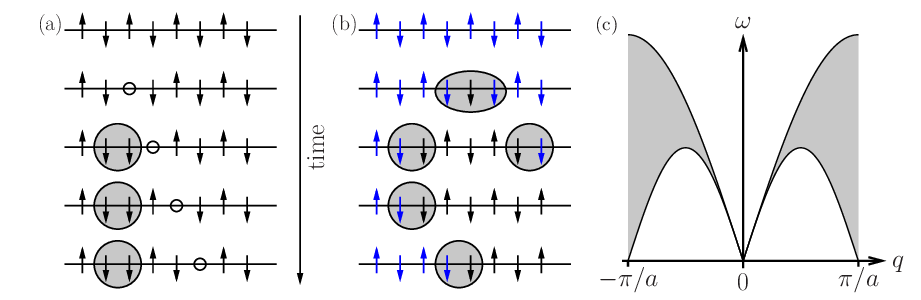}
    \caption{(a) Spin and charge separate in a
      one-dimensional electron system. A spin up electron is removed, leaving a
      hole behind. As
      the hole moves down the chain it leaves behind it a spin
      excitation (circled).
      (b) In a spin chain we create a $s=1$ magnon by flipping a spin. This
      is not a stable excitation but can split into a pair of $s=1/2$ spinons (circled)
      that can propagate via flips of spin pairs.
      (c) Spinons pairs give rise to a continuum of excitations in the
      chain (grey area) lying between the magnon dispersion (upper
      curve) and the single spinon dispersion (lower curve). 
      \label{fig6}}
  \end{center}  
\end{figure}

An important excitation found in most spin-liquid models is the
spinon. The spinon is an $s=1/2$, neutral fermion and therefore an example of
a fractional excitation.
The idea of  fracturing of the quantum numbers of an
electron is easiest to see in one dimension \cite{giamarchi}, which demonstrates the
remarkable feature that
the apparently fundamental properties of the electron: spin and
charge, can break into two. We imagine a set of electrons arranged along a
one-dimensional line as shown in Fig.~\ref{fig6}(a), with electron
spins aligned up-down-up-down, so the overall spin of the system is zero. 
If we remove an
electron from this system then we leave behind a hole, which can be thought of
as an excited state of charge (known as a holon in this context). If we move an electron along the line
into the empty space without changing its spin, then we see that the
hole is mobile. This has a consequence: it leaves two like spins as
neighbours forming a $s=1/2$ spinon excitation. As we slide
electrons around we see that the holon can move independently of the
spinon. Similarly by flipping a pair of spins we can move the spinon
around. The spin excitation and the hole excitation are independent:
we have spin-charge separation.

In the purely magnetic case (i.e.\ without the charge excitation), we
attempt to measure the dispersion by flipping a spin
[Fig.~\ref{fig6}(b)] to create a $s=1$ magnon excitation with
wavevector $q$. This is not
stable on the 1D chain and so falls apart into a pair of $s=1/2$
spinons that propagate via pairs of spin flips with wavevectors
$q_{1}$ and $q_{2}$, where $q=q_{1}+q_{2}$. We therefore expect to
measure (in an inelastic neutron scattering measurement) a
continuum of possible excitations [the grey shading in
Fig.~\ref{fig6}(c)] lying between the dispersion curves of a single
spinon and the single magnon. It is the characteristic continuum of
excitations that tells us we have fractional excitations. 


\subsection{Representing spins as fermions}

It is rather surprising that we can form spinons from the spins in a
magnet. In this section we discuss how this is possible within a
mean-field picture of magnetism.\footnote{Although our discussion is
  given in terms of fractionalising spins to form fermionic spinons,
  it is also possible to choose to form bosonic spinons via the use of
  the Schwinger boson representation of spin \cite{coleman}. Such an
  approach has the appealing feature that some states can be described
in terms of the Bose-Einstein condensation of such spinons~\cite{Xu}.}
The idea will be to fractionalise the spins into parts or ``partons''. If we
guess how to make the split correctly, the partons might represent the
true excitations of the system. As has been  commented \cite{wen,savary}, some practitioners are uncomfortable with
this approach, but we shall willingly suspend disbelief for the
moment.

Let's return to two dimensions and to  the Heisenberg antiferromagnet
with Hamiltonian
$\hat{H} = -\sum_{\langle
  ij\rangle}J_{ij}\hat{\boldsymbol{S}}_{i}\cdot\hat{\boldsymbol{S}}_{j}$
and make a {\it mean field approximation} \cite{lancaster}. This essentially causes a spin to
sit in the constant average magnetic field of its ordered neighbours, allowing
us to compute its dynamics. The usual recipe to do this replaces the original
Hamiltonian by
\begin{equation}
  \hat{H}_{\mathrm{mf}} = -\sum_{\langle ij\rangle}J_{ij}\left[
  \langle\hat{\boldsymbol{S}}_{i}\rangle\cdot\hat{\boldsymbol{S}}_{j}
  +
  \hat{\boldsymbol{S}}_{i}\cdot  \langle\hat{\boldsymbol{S}}_{j}\rangle
  -
   \langle\hat{\boldsymbol{S}}_{i}\rangle\cdot  \langle\hat{\boldsymbol{S}}_{j}\rangle\right].
\end{equation}
where $\langle \hat{\boldsymbol{S}}_{i}\rangle = \langle
\Phi_{\mathrm{mf}}|\hat{\boldsymbol{S}}_{i}|\Phi_{\mathrm{mf}}\rangle$
and $|\Phi_{\mathrm{mf}}\rangle$ is the mean field ground state of the
system. 
Although this approximation represents the spirit of this section,
this prescription won't actually work for a QSL,
since the states with which we're interested have $\langle
\boldsymbol{S}_{i}\rangle=0$, as we expect them to be disordered. 

To make progress, we perform a
transformation trick that
allows us to represent the spin as a chargeless fermion.\footnote{It's worth noting that such transformations have a rich history in many-body physics.
Jordan and Wigner spotted that a single spin state can be thought of
as an empty, or singly-occupied fermion state using a mapping
$|\uparrow\rangle \equiv f^{\dagger}|0\rangle$ and $|\downarrow\rangle \equiv
|0\rangle$. However, to deal with more than one spin they must add a
phase factor, called a string, to each fermion. \cite{coleman}}
We introduce spinon operators $\hat{f}_{i\alpha}$ where $i$ is a site
on the lattice and
$\alpha=1,2$ \cite{wen,zhou}. These are two-component, fermionic $s=1/2$, charge-neutral
operators  (sometimes called Abrikosov fermions).
There are two sorts of these ($f_{i1}$ spinons and $f_{i2}$ spinons). 
A spin operator is represented by spinons as
\begin{equation}
\hat{\boldsymbol{S}}_{i} =
\frac{1}{2}\hat{f}^{\dagger}_{i\alpha}\boldsymbol{\sigma}_{\alpha\beta}\hat{f}_{i\beta}
= \left(
  \begin{array}{cc}
\hat{f}^{\dagger}_{i1}& \hat{f}^{\dagger}_{i2}
  \end{array}\right)
\left(
  \begin{array}{cc}
    \boldsymbol{\sigma}_{11}&\boldsymbol{\sigma}_{12}\\
    \boldsymbol{\sigma}_{21}&\boldsymbol{\sigma}_{22}\\ 
  \end{array}
\right)
\left(
  \begin{array}{c}
    \hat{f}_{i1}\\
    \hat{f}_{i2}
  \end{array}
  \right).
\end{equation}
In these equations, sums over the Greek indices, which take the values
1 and 2, are implied.

The result of the spinon
transformation is that the magnetic Hamiltonian becomes
$
  \hat{H} = \frac{1}{2}\sum_{\langle ij \rangle}J_{ij}\hat{f}^{\dagger}_{i\alpha}
  \hat{f}_{j\alpha}\hat{f}^{\dagger}_{j\beta}\hat{f}_{i\beta} + \mathrm{const}.
$
To recap, we have recast the magnetic Hamiltonian in terms of
chargeless spinons which yields an expression in the spirit of the
tight-binding model of eqn~\ref{eqn:tb}, that can be interpreted in
terms of fermions hopping between sites.
By making this transformation we've allowed there to be four possible spinon
states per site ($|f_{i1}f_{i2}\rangle = |00\rangle$, $|01\rangle$,
$|10\rangle$, $|11\rangle$), where previously there were only two spin
states ($|\uparrow\rangle$ and $|\downarrow\rangle$).
To
make sure there's one fermion per site we should, strictly speaking, set the constraint
$\hat{f}^{\dagger}_{i1}\hat{f}_{i1} +
\hat{f}^{\dagger}_{i2}\hat{f}_{i2}=1$, which reduces the possible states
down to $|10\rangle\equiv |\uparrow\rangle$ and
$|01\rangle\equiv|\downarrow\rangle$ only.

We can now make
progress by imposing a mean-field approximation on the spinon model \cite{wen,zhou}.
Recall that this involves taking an average of combinations of
operators. 
This is
carried out here by
relaxing the constraint on the number of fermions per site, such that
we 
set
$\langle \hat{f}^{\dagger}_{i\alpha}\hat{f}_{i\alpha}\rangle=1$,  which
says that the {\it average} number of fermions per site is now one.
In
practice this is done using a Lagrange multiplier $V_{i}$, leading to a
new term
$V_{i}(f^{\dagger}_{i\alpha}f_{i\alpha}-1)$ in the Hamiltonian. We
enact the mean-field approximation, which now amounts to taking averages
of combinations of pairs of spinon operators, and throws up
the answer
\begin{equation}
  \hat{H}_{\mathrm{mf}} = \frac{1}{2}\sum_{\langle ij \rangle}J_{ij}
  \left[
    \hat{f}^{\dagger}_{i\alpha}\hat{f}_{j\alpha}\chi_{ji}
    + h.c.-|\chi_{ij}|^{2} +
    \sum_{i}V_{i}(\hat{f}^{\dagger}_{i\alpha}\hat{f}_{i\alpha}-1)
  \right],
\end{equation}
where $h.c.$ is the Hermitian conjugate of the preceding term. 
The new quantity $\chi_{ij} = \langle
\hat{f}^{\dagger}_{i\alpha}\hat{f}_{j\alpha}\rangle=
\langle
\hat{f}^{\dagger}_{i1}\hat{f}_{j1}\rangle
+
\langle
\hat{f}^{\dagger}_{i2}\hat{f}_{j2}\rangle
$ is key here. It measures the
amplitude for either sort of fermion to hop from site $j$ to site
$i$. 
The quantities $\chi_{ij}$ and $V_{i}$
don't change if we rotate
spins, so we can conclude that the ground state also has this
rotational invariance. The model can now yield spin-liquid
ground states. 

From the previous discussion, we would conclude that there is a
disordered, correlated ground state whose excitations are spinons. 
However, to get a reliable picture that maps to the Hilbert space of the
original spin model,
we must also include the possibility of low-energy fluctuations in $\chi_{ij}$ 
by saying
$\chi_{ij} = \bar{\chi}_{ij}{\rm e}^{-\mathrm{i}\theta_{ij}}$
and obtain
\begin{equation}
  \hat{H}_{\mathrm{mf}} = \frac{1}{2}\sum_{\langle ij \rangle}J_{ij}
  \left[
    \hat{f}^{\dagger}_{i\alpha}\hat{f}_{j\alpha}\bar{\chi}_{ji}{\rm e}^{-\mathrm{i}\theta_{ij}}
    + h.c. +
    \sum_{i}V_{i}(\hat{f}^{\dagger}_{i\alpha}\hat{f}_{i\alpha}-1),
  \right]
\end{equation}
We see that if we make the local phase transformation
$\theta_{ij}\rightarrow \theta_{ij} + \phi_{i}-\phi_{j}$ and
$\hat{f}_{i}\rightarrow \hat{f}_{i}{\rm e}^{\mathrm{i}\phi_{i}}$ the Hamiltonian
is unchanged. We have discovered the $U(1)$ gauge structure of the model. 
We identify the $V_{i}$ and $\theta_{ij}$, which interact with the spinons
in the same form, as components of a
gauge field,
and conclude that the
excitations of the theory are spinons {\it coupled} to this $U(1)$ gauge
field.
To make further progress we need to specify the form of $\chi_{ij}$,
which then allows us to derive the dispersion of the
excitations for specific QSL models. This can result in a range of states
including RVB-type models and a several $Z_{2}$ spin liquids \cite{wen}. 

For now, we can recap the main features of this approach to finding QSLs: in order to
use a mean-field theory to describe magnetically-disordered states we
have needed to fractionalise the spins before taking
the averages on which the mean-field technique relies. The result is a theory that predicts magnetically
disordered ground states with neutral $s=1/2$ fermion
excitations.
On allowing
fluctuations in the averaged quantities we obtain a gauge structure,
where we interpret these fluctuations quantum mechanically
as bosonic excitations in the gauge
field.
In short, we split up the spins at the start, and we use bosonic glue
(via the gauge-field component $V_{i}$)
to stick them back together!
This results in a Hamiltonian model that can support a spin liquid
ground state,
with predictions of its excitations. 

This general approach can be extended to more complicated gauge
structures to make a range of spin liquids states.
We expect stable mean-field QSLs to have a non-zero energy gap against gauge field
fluctuations.  The stable QSLs always contain neutral $s=1/2$
spinons with short range interactions between them.
The mean field approach predicts four families of spin liquid in (2+1)
dimensions, characterised by the nature of the energy gap between the
ground state and the spinon and gauge excitations. (See Ref.~\cite{wen}
for the full story.)

\noindent
{\it Rigid spin liquids}
are topologically ordered, which means both spinon excitations and the
excitations in the gauge field have non-zero energy gaps. The gaps
lead the QSLs to be stable since there are no low-energy excitations
to destabilise the states. 
 Examples include $Z_{2}$-gapped liquids and chiral liquids. 

\noindent
{\it Bose spin liquids}
are a class characterised by spinon excitations with an energy gap, and
gapless $U(1)$ gauge-boson excitations. These are not stable in (2+1)
dimensions. 

\noindent
{\it Fermi spin liquids}
have
gapless $s=1/2$ spinon excitations with short-range
interactions between them. The name derives from Fermi liquids, which
are characterised by  their gapless electronic excitations.
Examples include $Z_{2}$-linear-, $Z_{2}$-quadratic-, $Z_{2}$-gapless
liquids,
where names refer to features of the spinon dispersion. 

\noindent
{\it Algebraic spin liquids}
have gapless excitations, but these are neither free bosons nor free
fermions. Excitations are massless fermions coupled to $U(1)$ gauge field.
Examples include the $U(1)$-linear liquid.

If you've succeeded in suspending disbelief until now, disbelief is
surely flooding back! Should we trust any of these manipulations?
Arguably, the jury is still out, but the approach does suggest the
features of 
some of the possible states that might be out there for discovery.
Ultimately experiment will be the arbiter and we shall discuss the
status of experiments at the end of the article. For now we shall turn
to a different means of realising a QSL that, unlike the mean-field
picture, does not rely on approximations. 

\section{Kitaev model}\label{sec:kitaev}
For many of the formative years of the subject, the search for QSLs
was inspired by the successes such as the FQH fluid and BCS
superconductivity,
that suggested that identifying
wavefunctions, rather than Hamiltonians, was the way to discover new
states of matter. Although this approach has indeed been transformative, there was
also a great deal of value in the later identification of Hamiltonian
models of QSLs. 
A major step forward has been the formulation
of the Kitaev model \cite{kitaev}. This is not least because it is
solvable, based on a simple magnetic Hamiltonian that has a spin-liquid
ground state.
The solution, which involves the use of operators representing
Majorana fermions (a hypothetical fermion that is its own
antiparticle), looks potentially rather unfamiliar, but the result is less so: 
it leads to  a $Z_{2}$ gauge field and vortex excitations, just like
we saw in Section~\ref{sec:toric}. Even
better, in one limit of interactions the system gives us back the
toric code model exactly. 

\begin{figure}
  \begin{center}
    \epsfig{file=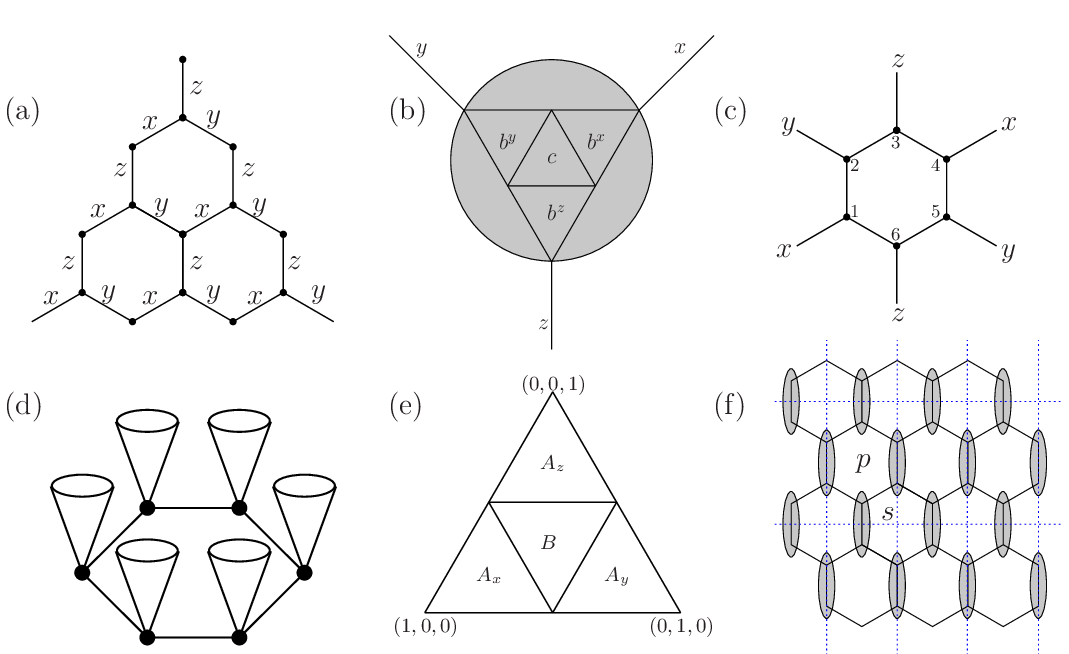,width=\columnwidth}\\
    \caption{The Kitaev model. (a) Honeycomb lattice with directional
      bonds.
      (b) To solve the model, we define a set
      of particle operators attached to each vertex of the
      lattice. (c) A plaquette in the lattice leads to the definition
      of the operator $\hat{W}_{p}$.
(d) Dispersion for $J_{x}=J_{y}=J_{z}$ showing Dirac cones. 
      (e) The triangular phase diagram 
      given in terms of $(J_{x},J_{y},J_{z})$.
(f) For $J_{z}\gg J_{x}, J_{y}$, we create superspins along the $z$
bonds and recreate the toric code lattice. A sample plaquette and star
is shown. 
      \label{fig:kitaev}}
\end{center}
\end{figure}

The Kitaev model is based on a planar honeycomb lattice with spins on each
vertex [Fig.~\ref{fig:kitaev}(a)].
The links are labelled $x$, $y$, $z$ and the Hamiltonian is written
\begin{equation}
  \hat{H} =
  -J_{x}\sum_{x}
  \hat{\sigma}^{x}_{i}\hat{\sigma}^{x}_{j}
-J_{y}\sum_{y}
\hat{\sigma}^{y}_{i}\hat{\sigma}^{y}_{j}
-J_{z}\sum_{z}
\hat{\sigma}^{z}_{i}\hat{\sigma}^{z}_{j}.
\end{equation}
To solve the model we introduce an operator that operates
on hexagonal plaquette $p$ shown in Fig.~\ref{fig:kitaev}(c) via
$  \hat{W}_{p} =
  \hat{\sigma}^{x}_{1}
  \hat{\sigma}^{y}_{2}
  \hat{\sigma}^{z}_{3}
  \hat{\sigma}^{x}_{4}
  \hat{\sigma}^{y}_{5}
  \hat{\sigma}^{z}_{6},
$
with eigenvalues $W_{p}=\pm 1 $. The $\hat{W}_{p}$  for each $p$
commutes
so we can describe the Hilbert space in terms of a set
 $\{W_{p}\}$. It will turn out that the ground states has $W_{p}=1$
 for all plaquettes.

 Although this is progress, the full solution requires swapping spins for a new set of
operators as we did in the last section.
Fascinatingly, these are the operators describing Majorana fermions.
Conventional fermions can be
created (with operator $\hat{d}^{\dagger}$) an annihilated ($\hat{d}$), with the
constraints $\hat{d}^{\dagger}\hat{d}^{\dagger}=0$ and
$\hat{d}\hat{d}=0$, and  anticommutator
$\hat{d}^{\dagger}\hat{d}+\hat{d}\hat{d}^{\dagger}=0$.
A Majorana fermion is its own antiparticle,  which means that
conjugating the charge of a Majorana
has no effect. In terms of operators this means we represent
Majoranas as real-valued.
Therefore, if we have two species of Majorana ($a$ and $b$ particles)
 then we could split up a fermion as $\hat{d}=(\hat{a}+\mathrm{i}\hat{b})/2$, where the
 Majorana operators for this representation are $\hat{a}$
 and $\hat{b}$. It follows that the Majoranas
 have the properties $\hat{a} = \hat{a}^{\dagger}$
and $\hat{b}=\hat{b}^{\dagger}$ (i.e.\ the creation and annihilation
operators are the same), 
$\hat{a}^{2}=\hat{b}^{2}=1$ (i.e.\ applying the operator twice has no effect) and
$\hat{a}\hat{b}+\hat{b}\hat{a}=0$ (the operators anticommute with each other). 

\subsection{The Kitaev model solution and its gauge structure}

For our problem \cite{kitaev}, we will need to introduce
four Majorana particles at each vertex of the lattice [Fig.~\ref{fig:kitaev}(b)]. 
At each vertex we therefore define a set of Majorana operators
$\hat{b}^{x}, \hat{b}^{y}, \hat{b}^{z}, \hat{c}$, and we write spin
operators in terms of these as
$\hat{\sigma}^{x} = \mathrm{i}\hat{b}^{x}\hat{c}$,
$\hat{\sigma}^{y} = \mathrm{i}\hat{b}^{y}\hat{c}$, and
$\hat{\sigma}^{z} = \mathrm{i}\hat{b}^{z}\hat{c}$.
The new operators square to unity and they
anticommute in the same way that the $\sigma$ matrices do. However, 
 we find
that 
we also  need an extra constraint:
identify operator $\hat{D} =
\hat{b}^{x}\hat{b}^{y}\hat{b}^{z}\hat{c}$  (which commutes with the spin operators) and, with the stipulation that
$D=\pm 1$, we guarantee a consistent
theory.\footnote{The point here is that the Hilbert space dimension for a single spin is 2 but for four majoranas it is 4, so we need to project out two spurious states per site.}
So why do we need four Majoranas? We'll give the $c$-field a
local $Z_{2}$ invariance, and the $b$-fields will be the gauge fields
that guarantee this.
This will give us a QSL ground state with
fermion excitations that derives from the $c$-field, along with
vortex excitations that will arise from the flux of the gauge field. 

With the proposed set of transformations, the Kitaev Hamiltonian becomes
\begin{equation}
  \hat{H} = \frac{\mathrm{\mathrm{i}}}{4}\sum_{jk}
 2J_{\alpha}\left(\mathrm{i}\hat{b}^{\alpha}_{j}\hat{b}^{\alpha}_{k}\right)
\hat{c}_{j}\hat{c}_{k}.
\end{equation}
The first thing to notice is that this looks like a tight-binding
model for the $c$-particles. 
Moreover, the shape of this Hamiltonian has the $c$-type Majoranas interacting
with an effective field $\mathrm{i}\hat{b}\hat{b}$. 
In this case, the combination
$\mathrm{i}\hat{b}^{\alpha}_{j}\hat{b}^{\alpha}_{k}=\hat{u}_{jk}$,
with $\alpha$ determined by the direction of the bond [i.e.\
$\alpha=\alpha(i,j)$], is a $Z_{2}$ gauge field,
since 
$u_{jk}=\pm
1$.\footnote{To see this,
note that the local transformation in this case would send
$c_{j} \rightarrow W_{j}c_{j}$ and then a change in the gauge field of
$u_{jk}\rightarrow W_{j}u_{jk}W^{-1}_{k}$ cancels out this effect.}
The operator $\hat{u}_{jk}$ commutes with the Hamiltonian, so we can split up the solutions
according to the values of $u_{jk}$ using the product around a plaquette $\hat{W}_{p} =\prod
\hat{u}_{jk}$, which we now see
evaluates the flux of the gauge field through a plaquette.  

The ground state has
$W_{p}=1$ for all plaquettes. This state
is a magnetically-disordered (i.e.\ spin-rotation invariant), spin-liquid phase. 
The Hamiltonian can then be diagonalised to reveal the dispersion for the
$c$-field excitations
of $
  E(\boldsymbol{k}) = \pm 2\left|
    J_{x}{\rm e}^{\mathrm{i}\boldsymbol{k}\cdot\boldsymbol{n}_{1}}
    +
        J_{y}{\rm e}^{\mathrm{i}\boldsymbol{k}\cdot\boldsymbol{n}_{2}}+J_{z}
  \right|,
$
where $\boldsymbol{n}_{1,2} = \left(\pm\sqrt{3}/2, 3/2\right)$. 
As usual, we characterise the material in terms of the gap against
excitations. An
example of the spectrum for $J_{x}=J_{y}=J_{z}$ is shown in
Fig.~\ref{fig:kitaev}(d) and has
several sets of excitations 
existing from $E=0$, each with a striking linear
dispersion. The points in $k$-space from which they emerge are known
as Dirac points, where the dispersion resembles that of highly
relativistic particles (i.e. with linear dispersion $E=|p|c$, where
$c$ is the speed of light). 

We can always find some fermion solution with $E=0$ as long as the exchange
constants obey triangle inequalities:
$|J_{x}|\leq |J_{y}|+|J_{z}|$ or $|J_{y}|\leq |J_{z}|+|J_{x}|$
or $|J_{z}|\leq |J_{x}|+|J_{y}|$. When these aren't obeyed, the
spectrum has a gap. 
We can draw a
triangular phase diagram of gapped and gapless phases  
 as shown in Fig.~\ref{fig:kitaev}(e). The phases $A_{i}$
are completely gapped,
while the phase $B$ is gapless. In addition to the  fermion excitations, we have $W_{p}=-1$
excitations in the gauge field. These are bosonic vortices.

Finally, as promised, in the limit of strong $J_{z}$-coupling we recover the toric code
Hamiltonian.
Physically, this is because the strong $z$-links tie the spins together to form effective
spins $|1\rangle = |\uparrow\uparrow\rangle$ and $|2\rangle =
|\downarrow \downarrow\rangle$ which decorate the edges of a square
lattice. This is shown in Fig.~\ref{fig:kitaev}(f), where a sample
plaquette and star are indicated.

\section{Realising the a QSL in materials}
We've seen several ways of describing different putative QSLs, but the question
remains whether the state is realised in real materials. As we said at
the start of this article, there is a wide-ranging literature claiming
QSL behaviour in a large number of materials. However, it's
probably fair to say that we do not have a case where a consensus has
been reached in favour of the occurrence of the QSL state. 
Despite this we have reason to be hopeful. We have dealt with (2+1)
dimensions in our discussion so far, but the analogous
one-dimensional (1D) case of the {\it spin-Luttinger liquid} is agreed
to be realised in some spin-chain materials, in which direct measurements of
spinons have been made, most directly using inelastic neutron
scattering (INS). The spin-Luttinger regime is characterised by a lack of
long-range magnetic order (LRO), algebraically-decaying spin correlations
and, most vividly, a continuum of spinon excitations [Fig.~\ref{fig6}(c)].
Here it is
unwise to say a material {\it is} a spin-Luttinger liquid, rather that
it behaves as a spin-Luttinger liquid under certain
conditions.
The 1D chain will be characterised by an intrachain interaction $J$. If the
chains have some small interchain interaction $J'$ then the material might well order
at a very low temperature $T_{\mathrm{N}}$ determined by $J'$. In order to observe 1D
physics we seek to be in a regime that $T_{\mathrm{N}}\ll T \ll
J$. This means we stay away from magnetic order, but the temperature
is low enough that the collective behaviour is promoted by $J$. (For
$T\gg J$ excitations will start to resemble single uncorrelated spin flips.)

A QSL will be realised in more than one dimension. Although we have
dealt with 2D in our discussion, there are also proposals for 3D QSL
states. 
To realise a QSL we seek to suppress magnetic order in
materials with strong spin-spin interactions. 
Low dimensionality is good for this, although many 2D materials enjoy
enough coupling in the third dimension to promote order at low
temperature. 
Magnetic order can also be suppressed by frustrating magnetic interactions. 
(When this occurs, small interactions can tip the system over into LRO
at a low-enough $T$, so these must be borne in mind.)
We also want to
promote quantum fluctuations and for this $s=1/2$ spins are most effective.
Other possibilities involve promoting higher-order exchange
interactions, such as next-nearest neighbour couplings. 

The experimental identification of a QSL is a knotty problem, as
discussed in Refs~\cite{knolle,matjaz2}. Clearly
we seek a system without magnetic order down to low temperature. A lack of order can be
difficult to establish, but sensitive local probes such as NMR and muon-spin
spectroscopy ($\mu^{+}$SR) are often used to search for weak signs of
order, or spin dynamics characteristic of disordered spin-liquid states.
We have stressed that the excitation spectrum is key to the
characterisation of a QSL system. Perhaps the most important question
is whether there is a gap against excitations and there are several ways
of investigating this that are sensitive to the density of states,
including thermodynamic, scattering and spectroscopic probes. 
Beyond this, a determination of the excitation spectrum might be
expected to supply the smoking gun evidence. INS remains a key
method here, but requires samples that are relatively large
compared to those required for other measurements. There is also the
problem that some nuclei have large neutron-capture cross sections, or
the material
contains hydrogen or other nuclei that cause a large degree of incoherent
scattering. Nevertheless, we look for
fractionalised quasiparticles which, like spinons in 1D, give broad responses in
scattering as they correspond to the creation of several particles. In
the absence of  momentum-resolved scattering measurements, there are predictions
for the temperature and magnetic field dependence of properties such as
heat capacity, spin relaxation or transport. In the case of the
latter, although QSLs should strictly be insulators, there are
possibilities for the state to couple to charges, allowing an
additional handle to probe the system. 

We have stressed that the key to the QSL problem is the special
pattern of quantum entanglements that underlie the physics. Finding a
means of evaluating these could prove the key to this subject.
The entanglement entropy $S$ of the many-particle ground state is useful here. For a
gapped system it follows the form $S=cL-\Gamma+...$,
where $L$ in the first term is the size of the system, while the second term quantifies long-range entanglements \cite{knolle}. This latter
part is only non-zero in a topological phase (for a $Z_{2}$ QSL,
$\Gamma=\ln 2$). The bad news, however, is that there isn't yet a
known way of measuring
 this quantity \cite{knolle}.

Below we briefly introduce four different candidate systems that have been
much discussed as possible approximate realisations of a QSL state.
Further discussion, including the many relevant references, can be
found in the detailed reviews of Refs.~\cite{savary,senthil,zhou,clark,wen2}.

{\bf Herbertsmithite} or ZnCu$_{3}$(OH)$_{6}$Cl$_{2}$ is a mineral,
originally discovered in a Chilean mine,
but since synthesised in the laboratory.
It comprises $s=1/2$ spins on a frustrated 2D kagome lattice interacting
with an exchange of $J=180$~K. The kagome lattice is formed from
corner sharing triangles (as distinct from edge-sharing triangles of
the triangular lattice), resulting in a larger degeneracy and
potential for realising QSL states.
No evidence of long-range magnetic order has been found in
Herbertsmithite down to 50~mK. Specific heat and some NMR measurements suggest gapless
behaviour, although a spin gap has also been identified on the basis
of Knight-shift measurements. 
Large crystals have enabled neutron scattering, but the broad
continuum of magnetic excitations found was not one predicted by spin liquid
theory. 
In fact,
the low-energy excitation spectrum appears dominated by impurity spins
that
originate from sites between the kagome planes. Although these sites should
host non-magnetic Zn$^{2+}$, they appear to contain enough magnetic Cu$^{2+}$ to
form a significant population of orphaned impurity moments (estimated
to be 5-10\%).
This raises the question of how robust a QSL might be with respect to
disorder, which is a question that is pertinent to all of the cases
discussed here. The presence of perturbing anisotropic
(Dzyaloshinskii-Moriya) interactions between intrinsic kagome spins
has also been found using electron spin resonance, which could have a
large effect on the ground state \cite{zorko}.

{\bf $\kappa$-(ET)$_{2}$Cu$_{2}$(CN)$_{3}$} is formed from a layered structure built from structural
dimers of planar BEDT-TTF molecules, sandwiched between layers of Cu$_{2}$(CN)$_{3}$.
Each dimer gives rise to a magnetic $s=1/2$
spin that form a 2D triangular lattice with a large exchange
$J>200$~K. This material is part of a family of
materials that show superconductivity and magnetic order, along with a
number of spin-liquid candidates. 
For $\kappa$-(ET)$_{2}$Cu$_{2}$(CN)$_{3}$ there is no evidence of
magnetic order from $\mu$SR or NMR.
The combination of constant low-temperature magnetic susceptibility $\chi$,
linear specific heat (i.e.\ like a metal $C=\gamma T$) and
a resulting Fermi-gas like Wilson ratio was suggestive of a gapless
phase with spin-carrying excitations. However conductivity and $\mu$SR
both suggested the presence of an energy gap. Owing to the tunability
of molecular materials, several other systems based on similar
building blocks also show some promise.

{\bf $R_{2}M_{2}$O$_{7}$ ($R=$rare earth, $M=$transition metal such as
  Ti or Sn)} are a well know frustrated series of 3D materials based
on spins in a 3D arrangement of corner-sharing tetrahedra, called the
pyrochlore lattice, which is known to exhibit a high degree of frustration.
Strong spin-orbit interactions and crystal field splitting can result
in systems described by an effective $s= 1/2$ theory.
Some of these materials are well-described by classical theories (e.g.\
Ho$_{2}$Ti$_{2}$O$_{7}$
and Dy$_{2}$Ti$_{2}$O$_{7}$). For example, in
Dy$_2$Ti$_2$O$_7$ 
the crystal-field anisotropy constrains the magnetic moments to lie
along the local $\langle 111 \rangle$ axes (i.e.\ directly in our out of the tetrahedron) and there is 
an effective local ferromagnetic coupling between these moments, as well as long-range dipolar couplings, which turn out to be important. 
As a result of this combination of interactions and the local
anisotropy, below about 1~K the system settles into a disordered {\it
  spin-ice state}, which is often described as a classical spin liquid
(i.e.\ a highly correlated magnetic system that avoids magnetic
order). Spin ice is characterised by a `2-in 2-out' spin configuration
(meaning that two spins point in  and two spins point out of each
tetrahedron), analogous to proton displacement vectors in Pauling’s
model of hydrogen disorder in water ice. 
The excitations in spin ice are created by reversing a single spin,
which produces a pair of effective magnetic monopoles which can move
independently through the lattice and interact with an emergent gauge field, but remain connected by a topological Dirac string of flipped spins between them. 
In contrast, strong quantum effects have been found in
$R_{2}$Ti$_{2}$O$_{7}$ with $R=$Yb, Er and Tb and in Pr$_{2}$Zr$_{2}$O$_{7}$. 
Of these Yb is the most studied, having been suggested to be
a quantum version of spin ice which features frustrated Ising
interaction. Such a model is thought to support a $U(1)$ QSL in some limits of its
parameters. 
 This picture has some experimental support (e.g.\ it features a
``pinch point'' feature in its neutron spectrum, predicted for
classical spin ice owing to dipolar spin correlations).
However, there is also good evidence for a  phase transition to a ferromagnetic ground state in high-quality
samples, and the unusual neutron spectrum has been suggested to arise
from an interplay of types of conventional order \cite{scheie}. 
This is another case where there is a degree of sample dependence, suggesting a role
for disorder. 

Finally, there are several candidate materials that might be
described by the {\bf Kitaev} Hamiltonian. 
Although the Kitaev Hamiltonian, with bond-dependent Ising interactions, appears somewhat contrived, it was suggested that partially-filled $t_{2g}$ levels in an
octahedral environment with strong
spin-orbit coupling could realise the model. 
The idea was originally to look at materials containing Ir$^{4+}$, which has an
effective $j=1/2$ moment, owing to the
orbital degeneracy being lifted 
in favour of a Kramers doublet. 
Studies of honeycomb $\alpha$-Na$_{2}$IrO$_{3}$ and
$\alpha$-Li$_{2}$IrO$_{3}$ revealed magnetic order, but also indicated that
the bond-directed Kitaev interactions were, to some extent, realised. 
Iridium is well
known for its high neutron capture cross section, making INS
difficult and so the
most-studied Kitaev candidate is not an Ir-hosting material, but
$\alpha$-RuCl$_{3}$.
Here $j=1/2$ ruthenium ions  form a honeycomb
lattice between Cl planes.
This system also orders ($T_{\mathrm{N}}=15$~K), and seems to host
conventional Heisenberg interactions in addition to bond-directed
ones.
However, there is evidence from
Raman scattering for a fermionic continuum and magnetic order seems to
be fully suppressed by a strong applied field, suggesting that the interactions
might be tuned to QSL behaviour. As a result, the system is sometimes
called a proximate spin liquid. 
So is the honeycomb lattice really sufficient for the Kitaev model? It
is possible that the spin liquid regime might only be stable in a tiny
region and the interpretation of data on these materials has been
hotly debated. Owing to the large amount of work on this subject at
the time of writing, this remains an area to watch.  

None of these systems uncontroversially show QSL behaviour, but all
have features suggesting they might be close. As with the 1D materials
the key to observation of this behavior will be looking at the realm
of applicability of the models to the materials. If the QSL states are
realised in some form, and it seems likely that they are, evidence
will likely continue to accumulate slowly as the vast parameter space
of possible materials is surveyed. One essential ingredient to
consider is disorder. This is an inevitable feature of condensed
matter systems and will be of relevance here. It is perhaps worth
keeping in mind its dual role in the related field of quantum Hall
physics: the FQH fluid was only observed in relatively clean samples,
but disorder is also believed to be necessary in order to observe the Hall
plateaux. It is possible disorder might adopt a similar dual role in
the story of the QSL, or at least resolve some of the contradictory
experimental evidence that is found in the current literature.

\section{Conclusion}
The study of quantum magnetism started with doubts about the existence
of the antiferromagnet but, through careful experiment and thoughtful
theoretical description, the field now boasts a wealth of established results that demonstrate
the fractionalisation of excitations in 1D and the existence of
topological excitations related to vortices. Advances in materials
preparation, including those towards the ability to engineer magnets from tuneable
building blocks, also provide hope that hitherto purely-theoretical models
might be realised in material systems. The quest to realise a quantum spin
liquid remains a much sought-after goal that goes well beyond simply
finding a material that fails to order at low temperatures. The aim of
this article has been to give a sense of what is at stake here beyond
a lack of magnetic order: namely a macroscopic manifestation of
a very specific type of quantum entanglement that presents us with a
zoology of elementary excitations over and above those that fit comfortably
within Landau's ``Standard Model'' of condensed matter physics. From this
point of view, the investigation of QSLs could help us in the next
phase of  determining the
ordering principles that quantum mechanics imposes on our material
world. 

\section*{Acknowledgements}
I am grateful the following colleagues for their many helpful comments and corrections:
 Stephen Blundell, 
 Theo Breeze, 
 Claudio Castelnovo, 
 Matja\v{z} Gomil\v{s}ek, 
 Thomas Hicken, 
 Ben Huddart,  
 Samuel Ladd and
 Francis Pratt. Particular thanks are due to
 John Chalker who made detailed comments on two early drafts of this
 article.


\section*{Biography}
Tom Lancaster is Professor of Physics at Durham University. His
research focuses on the physics of magnetism in one and
two dimensions along, more recently, with the physics of topological objects in
magnetism. 
He is the coauthor, with Stephen Blundell, of Quantum Field Theory for
the Gifted Amateur.


\end{document}